\documentclass[a4paper,11pt]{article}

\usepackage{jheppub}

\usepackage{graphicx}
\usepackage{dcolumn}
\usepackage{bm}
\usepackage{float}
\usepackage{subfig}
\usepackage{slashed}
\usepackage{color}
\usepackage{amsmath}
\usepackage[section]{placeins}
\usepackage{braket}
\usepackage{upgreek}
\usepackage[normalem]{ulem}
\usepackage[bottom]{footmisc}
\usepackage{tikz}
\usepackage{xcolor}
\usetikzlibrary{arrows.meta,positioning,fit,calc}

\definecolor{jetblue}{HTML}{E4EBF2}
\definecolor{jetblueedge}{HTML}{526D86}

\definecolor{convteal}{HTML}{D8EDE7}
\definecolor{convtealedge}{HTML}{287567}

\definecolor{poolfill}{HTML}{E5F0D8}
\definecolor{pooledge}{HTML}{658744}

\definecolor{headviolet}{HTML}{EAE3F3}
\definecolor{headvioletedge}{HTML}{79639A}

\definecolor{quarkfill}{HTML}{F7E6BF}
\definecolor{quarkedge}{HTML}{9B6A13}

\definecolor{gluonfill}{HTML}{F3DFE2}
\definecolor{gluonedge}{HTML}{A65361}

\definecolor{ink}{HTML}{263746}

\newcommand{\pT}{p_T}
\newcommand{\ptjet}{p_T^{\rm jet}}

\newcommand{\LN}{L_N}
\newcommand{\Mg}{M}

\title{Dissecting flavor tagging with graph neural networks}

\author[a,b]{Ankita Budhraja,}
\author[c]{Sunny Kumar Singh}
\author[a]{and Balbeer Singh}
\affiliation[a]{LIP, Av.\ Prof.\ Gama Pinto, 2, P-1649-003 Lisboa, Portugal}
\affiliation[b]{Nikhef, Theory Group, Science Park 105, 1098 XG, Amsterdam, The Netherlands}
\affiliation[c]{Department of Physics, Indian Institute of Technology Bombay, Mumbai 400076, India}
\emailAdd{abudhraja@lip.pt}
\emailAdd{balbeers@lip.pt}
\emailAdd{sunnysingh541@gmail.com}

\abstract{Graph neural networks such as ParticleNet are widely used for quark/gluon jet tagging. We study how the discriminating power of such a network is distributed between constituent multiplicity and the remaining structure of the jet, differentially in jet transverse momentum and multiplicity. To this end, we train ParticleNet on quark and gluon jets from $Z$+jet production in \textsc{Pythia}, with the two classes matched in transverse momentum and rapidity over the wide range $80\le\ptjet\le400$~GeV. We find that the network learns how the fraction of gluon jets changes with the multiplicity in every $\ptjet$ bin, including the shift of the multiplicity at which quark and gluon jets are equally likely, which implies that it uses the jet scale (reconstructed from its constituents) jointly with its substructure. At fixed multiplicity, the gluon probabilities it assigns agree on average with the observed gluon fractions, but are widely spread about the mean. This allows us to interpret its output as the sum of the multiplicity log-likelihood ratio and a residual log-likelihood ratio at fixed multiplicity, which allows the network to correctly identify $20$--$34\%$ of the jets whose multiplicity favors the opposite class. By rescaling the radial spread of the constituents at fixed multiplicity, we find that the network considers broader jets more gluon-like, and that this response is largest at low and intermediate multiplicity, where the multiplicity alone does not decide the class of a jet.}


\begin{document}
\maketitle

\section{Introduction}
\label{sec:intro}

Machine learning (ML) has emerged as a powerful tool for particle identification, tagging and energy estimation in high-energy collision experiments at facilities such as the Large Hadron Collider (LHC), addressing the limitations of traditional selection criteria in capturing complex and high-dimensional correlations in data~\cite{Larkoski:2017jix,Plehn:2022ftl, Guest:2018yhq,Feickert:2021ajf}. This is because ML approaches learn discriminating patterns and features directly from the input data, leading to significant improvements in classification accuracy and energy resolution~\cite{Larkoski:2017jix}.

Jets are collimated sprays of hadrons that arise when energetic quarks and gluons produced in a hard scattering undergo successive branchings, generically called parton showering, until their virtuality reaches $\mathcal{O}(1~{\rm GeV})$, where they confine into color-neutral hadrons~\cite{Salam:2010nqg}. Identifying whether a jet was initiated by a quark or by a gluon is one of the most persistent classification problems in collider physics~\cite{Gallicchio:2011xq,Gallicchio:2012ez,Larkoski:2014pca,Gras:2017jty}. It enters new-physics searches, where quark and gluon jets appear as background, and it provides a useful handle on the color-charge dependence of jet quenching in heavy-ion collisions~\cite{Brewer:2020och,Apolinario:2021olp,Lai:2021ckt}. Since partons are not observed directly, beyond leading order the quark or gluon origin of a jet can only be defined operationally~\cite{Komiske:2018vkc,Metodiev:2018ftz}, and a jet is identified as ``quark-like'' or ``gluon-like'' through its substructure. The larger color charge of the gluon, $C_A/C_F=9/4$, implies that a gluon jet typically radiates more, tend to contain more constituents, are broader, and share their momentum more evenly among the constituents. Standard jet substructure observables such as constituent multiplicity, momentum dispersion, jet angularities and jet charge are designed to exploit such differences in the underlying quantum chromodynamics (QCD) emission patterns of quarks and gluons~\cite{Gallicchio:2012ez,Larkoski:2014pca,Krohn:2012fg,Frye:2017yrw}. However, in most scenarios the constituent multiplicity turns out to be among the strongest single discriminant, although it is not infrared and collinear safe and its discriminating power differs considerably between hadronization models~\cite{Gras:2017jty,Frye:2017yrw,ATLAS:2014vax}.

Over the past decade, machine-learning methods have been extensively used to go beyond these traditional observables and exploit the full substructure of the jet~\cite{Komiske:2016rsd,Kasieczka:2018lwf,Komiske:2018cqr,Dreyer:2021hhr,Romero:2021qlf,Athanasakos:2023fhq,Bright-Thonney:2022xkx,Cogan:2014oua,deOliveira:2015xxd,Komiske:2017aww,Kasieczka:2019dbj}. More recently, graph neural networks (GNNs) that treat a jet as a dynamically constructed graph over its constituents have been established to provide remarkable improvements over previously employed ML architectures~\cite{Shlomi:2020gdn,Dreyer:2020brq}. By treating the jet as a graph and retaining the information about the four-momenta of the particles in the jet, GNN-based architectures can capture relational structures between jet constituents that otherwise remain elusive. Among these, ParticleNet~\cite{Qu:2019gqs}, which builds a $k$-nearest-neighbor graph over jet constituents and updates it through EdgeConv operations~\cite{Wang:2018dgcnn}, has become a benchmark architecture for jet-tagging studies, with subsequent advances using transformers~\cite{Qu:2022mxj} and quantum algorithms~\cite{Jobilal:2026qgnn} to further push classification performance. Taggers of this type are now also actively used in experimental analyses, see for e.g.~\cite{ATLAS:2025depart}.

Recently, attempts towards understanding what these networks learn have been made through relevance propagation, Shapley values, their latent representations, symbolic regression and mechanistic analyses~\cite{Agarwal:2020fpt,Khot:2022aky,Grojean:2022mef,Mokhtar:2022pwm,Vent:2025ddm,Patel:2026xai,Rai:2026mech,Jobilal:2026qgnn,Bach:2015lrp,Shapley:1953,Lundberg:2017shap,Cranmer:2023pysr}. For quark/gluon tagging, these studies have shown that the multiplicity accounts for a large part of the discriminating power of such networks, and that a small set of additional observables, related to the radial energy flow and the fragmentation pattern, recovers much of the remaining discrimination~\cite{Vent:2025ddm}. Similarly, for other jet-classification tasks, the decisions of a network have been mapped onto a small set of interpretable jet observables~\cite{Faucett:2020giy}. However, these results characterize the tagger on average over all jets. It remains to be understood how the information beyond the multiplicity depends on the jet itself: for which multiplicities and transverse momenta it matters most, and how often it is strong enough to correctly identify jets whose multiplicity is more typical of the opposite class.

In this paper, we address these questions for a ParticleNet classifier trained on quark and gluon jets matched in transverse momentum and rapidity over $80\le\ptjet\le400$~GeV. We first quantify the classification achieved through standard QCD jet observables and study the gain obtained through a simultaneous measurement of these observables. To this end, we construct boosted decision tree (BDT) classifiers from different sets of observables, and find that multiplicity, momentum dispersion and angularities carry largely shared information. While the momentum dispersion alone is nearly as discriminating as multiplicity at low $\ptjet$, multiplicity becomes increasingly dominant in discrimination towards high $\ptjet$. 

We then turn to the network itself and test whether its output can be read as the probability that a jet is a gluon jet, first for jets of fixed multiplicity. This allows us to write the network output as the sum of the multiplicity log-likelihood ratio and a residual log-likelihood ratio at fixed multiplicity, and to quantify how often the residual term is strong enough to correctly identify jets whose multiplicity favors the opposite class. By selecting for the jets whose multiplicity favors the opposite class, we further perform controlled deformations of the jet constituents, i.e. by rescaling the radial spread of the jet and study how the network responds to the width of the jet as a function of its multiplicity and transverse momentum, finding this contribution to dominate at  low and intermediate multiplicity.

The remainder of this paper is organized as follows. In section~\ref{sec:gnn}, we discuss the graph neural network architecture we employ, and in section~\ref{sec:setup} we provide details of the generation of the event samples, the kinematic matching of quark and gluon jets and the training. Section~\ref{sec:reference} discusses quark/gluon discrimination with the multiplicity likelihood ratio and with global jet observables. In section~\ref{sec:perf}, we present the quark vs.\ gluon performance of the network, investigate the substructure of the jets as classified by the network and provide a probabilistic interpretation of the network output. Finally, in section~\ref{sec:beyond}, we separate the multiplicity term from a residual log-likelihood ratio at fixed multiplicity and study the response of the network to the radial spread of the jet. We conclude in section~\ref{sec:summary}.

\section{Graph neural networks}
\label{sec:gnn}

Graph neural networks (GNNs) are a class of deep-learning models designed to operate on graph-structured data, learning node representations by iteratively aggregating and updating information from each node's neighbors. This message-passing mechanism allows the network to capture the relational structure encoded in the graph instead of assuming a fixed sequence as in conventional architectures~\cite{Gilmer:2017mpnn}. Following the conventional GNN construction, we represent each particle inside the jet as a node, with the particle's kinematics and identity as node features, as summarized in table~\ref{tab:graph_representation}. The edges are defined by connecting each particle to its $k$ nearest neighbors, encoding local particle-to-particle proximity within the jet. In the first block, the nearest neighbors are determined by the angular distance
\begin{equation}
\Delta R_{ij}=\sqrt{\Delta\eta_{ij}^2+\Delta\phi_{ij}^2}\,,\qquad \Delta\eta_{ij}=\eta_i-\eta_j\,,\quad \Delta\phi_{ij}=\phi_i-\phi_j\,,
\label{eq:dR}
\end{equation}
between particles $i$ and $j$.

\begin{table}[t]
\centering
\begin{tabular}{|p{3cm}|p{12cm}|}
\hline
\textbf{Graph Element} & \textbf{Representation in a Jet} \\
\hline
Graph & A single jet, represented as a point cloud of its constituent particles, $G = (V, E)$ \\
\hline
Nodes & Each particle within the jet is one node in the graph, $v_i \in V,\ i = 1, \dots, N$ \\
\hline
Node Features & Each node carries the feature vector of that particle, $\mathbf{x}_i=\big(\Delta\eta_i,\Delta\phi_i,\Delta R_i,\ln p_{T,i},\ln E_i,\ln \tfrac{p_{T,i}}{p_{T,{\rm jet}}},\ln\tfrac{E_i}{E^{\rm jet}},q_i,\text{PID}_i\big)$, with $\Delta\eta_i,\Delta\phi_i,\Delta R_i$ measured relative to the jet axis. \\
\hline
Edges & Each node is connected to its $k$-nearest neighbours $e_{ij} = (v_i, v_j),\ v_j \in \mathcal{N}_k(v_i)$ \\
\hline
\end{tabular}
\caption{Graph representation of a single jet in a given collision. Here $V,E$ denotes number of nodes and pair-wise edges (connections) between nodes. The notation $e_{ij}$ represents edge connections between nodes. The node features $q_i$ represent their electric charge and PID$_i$ is the particle type of the constituent $i$, encoded by five indicators (electron, muon, charged hadron, neutral hadron, photon), such that each node carries 13 features.}
\label{tab:graph_representation}
\end{table}

Having the graph representation of a jet, the core operation we perform to update the node features is EdgeConv~\cite{Wang:2018dgcnn}, which for each node $v_i$ computes an edge feature
\begin{equation}
e_{ij}=h_\Theta(\mathbf{x}_i,\,\mathbf{x}_j-\mathbf{x}_i)\qquad \text{for every neighbor } \mathbf{x}_j\in\mathcal{N}_k(v_i)\,,
\end{equation}
where $h_\Theta$ is a learnable function that combines the node features with the relative difference to its neighbor. This allows the network to learn individual particle kinematics as well as the local relational structure within the jet. The updated feature of node $v_i$ is then obtained by aggregating over its neighborhood via a channel-wise symmetric aggregation, which in ParticleNet is the mean~\cite{Qu:2019gqs},
\begin{equation}
\mathbf{x}_i'=\frac{1}{k}\sum_{j\in\mathcal{N}_k(v_i)} e_{ij}\,,
\label{eq:edgeconv}
\end{equation}
which ensures that the operation is invariant under the ordering of the particles.

Several such EdgeConv blocks are stacked sequentially and, crucially, the $k$-nearest-neighbor graph is recomputed after each block using the updated feature space rather than the original $(\Delta\eta,\Delta\phi)$ coordinates, so the network dynamically redefines particle ``closeness'' as it learns increasingly abstract representations of the jet substructure. The final node features are then aggregated via global average pooling and passed through fully connected layers to produce the jet-level classification output, which in our case is quark vs.\ gluon.

\begin{figure}[t]
	\centering
	\begin{tikzpicture}[
  font=\sffamily\scriptsize,
  >={Stealth[length=1.5mm]},
  flow/.style={-{Stealth[length=1.5mm]},draw=ink,line width=0.6pt},
  box/.style={draw=ink!70,line width=0.6pt,rounded corners=2pt,
              align=center,inner sep=3pt},
  stage/.style={font=\sffamily\bfseries\scriptsize,text=ink},
  note/.style={font=\sffamily\tiny,text=ink,align=center}
]

\node[box,fill=jetblue,draw=jetblueedge,text width=2.7cm,minimum height=1.12cm]
  (input) at (0,0) {
    \textbf{Jet as a graph}\\[2pt]
    Particles $v_i$ with features\\[2pt]
    $\Delta\eta_i,\Delta\phi_i,\Delta R_i$\\
    $\ln p_{T,i},\ln E_i,q_i,\mathrm{PID}_i$
  };
\node[stage] at (0,1.55) {Input};

\draw[draw=convtealedge,fill=convteal!10,rounded corners=3pt,line width=0.65pt]
  (2.00,1.02) rectangle (5.28,-1.02);

\node[box,fill=convteal,draw=convtealedge,text width=2.52cm,minimum height=0.52cm]
  (edge1) at (3.64,0.68) {\textbf{EdgeConv 1}\\[-1pt]\scriptsize Channels: $(64,64,64)$};
\node[box,fill=convteal,draw=convtealedge,text width=2.52cm,minimum height=0.52cm]
  (edge2) at (3.64,0) {\textbf{EdgeConv 2}\\[-1pt]\scriptsize Channels: $(128,128,128)$};
\node[box,fill=convteal,draw=convtealedge,text width=2.52cm,minimum height=0.52cm]
  (edge3) at (3.64,-0.68) {\textbf{EdgeConv 3}\\[-1pt]\scriptsize Channels: $(256,256,256)$};
\node[stage] at (3.64,1.55) {Graph convolutions};

\node[note,text width=3.45cm] at (3.64,-1.85) {
  Mean aggregation; $k=16$ nearest neighbours\\
  $k$-NN graph recomputed after each block
};

\node[box,fill=poolfill,draw=pooledge,text width=1.58cm,minimum height=0.98cm]
  (pool) at (6.78,0) {
    \textbf{Global average}\\
    \textbf{pooling}\\[2pt]
    $\displaystyle \frac{1}{N}\sum_i\mathbf{x}_i$
  };
\node[stage] at (6.78,1.55) {Jet representation};

\node[box,fill=headviolet,draw=headvioletedge,text width=1.48cm,minimum height=0.52cm]
  (fc) at (9.30,0.39) {\textbf{Dense 256}\\ReLU};
\node[box,fill=headviolet,draw=headvioletedge,text width=1.48cm,minimum height=0.52cm]
  (drop) at (9.30,-0.39) {\textbf{Dropout}\\$0.1$};
\node[box,fill=headviolet,draw=headvioletedge,text width=1.50cm,minimum height=0.62cm]
  (softmax) at (11.45,0) {\textbf{Dense 2}\\Softmax};
\node[stage] at (10.38,1.55) {Classification head};

\node[box,fill=quarkfill,draw=quarkedge,text width=1.30cm,minimum height=0.58cm]
  (qout) at (13.62,0.39) {\textbf{Quark jet}\\$P(q\mid\mathrm{jet})$};
\node[box,fill=gluonfill,draw=gluonedge,text width=1.30cm,minimum height=0.58cm]
  (gout) at (13.62,-0.39) {\textbf{Gluon jet}\\$P(g\mid\mathrm{jet})$};
\node[stage] at (13.62,1.55) {Output};

\draw[flow] (input.east) -- (2.00,0);
\draw[flow] (5.28,0) -- (pool.west);
\draw[flow] (pool.east) -- ++(0.20,0) |- (fc.west);
\draw[flow] (fc.south) -- (drop.north);
\draw[flow] (drop.east) -- (softmax.west);
\draw[flow] (softmax.east) -- ++(0.20,0) |- (qout.west);
\draw[flow] (softmax.east) -- ++(0.20,0) |- (gout.west);

\end{tikzpicture}
	\caption{Schematic of the ParticleNet architecture used in this work. }
	\label{fig:network}
\end{figure}
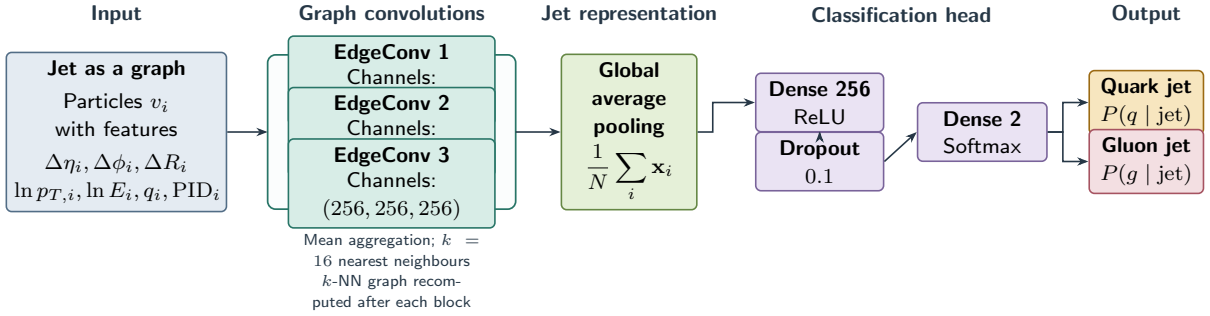

We use the ParticleNet architecture of ref.~\cite{Qu:2019gqs}. The network is built from a sequence of three EdgeConv blocks, each with $k=16$ nearest neighbors,
and with output channel dimensions $(64,64,64)$, $(128,128,128)$ and $(256,256,256)$ for the successive blocks. Within each block, the edge function $h_\Theta$ is implemented as a three-layer multi-layer perceptron, with each layer followed by batch normalization and a ReLU activation, allowing the network to learn a non-linear combination of a particle's own features and its relative separation from each neighbor~\cite{Ioffe:2015batchnorm}. Each block contains a shortcut connection, which stabilizes training and allows information from earlier, more physically interpretable feature spaces to propagate alongside the increasingly abstract learned representations~\cite{He:2016resnet}. The output of the last block is aggregated via global average pooling into a single jet-level representation; in contrast to ref.~\cite{Qu:2019gqs}, the outputs of the three blocks are not concatenated before pooling, and the input features are not normalized by a batch-normalization layer. This is passed through a fully connected layer with 256 units and dropout of 0.1 to a two-unit output layer with a softmax activation~\cite{Srivastava:2014dropout}, giving the predicted probability of the jet being quark- or gluon-initiated. A schematic representation of the \textsc{ParticleNet} architecture we utilize is shown in fig.~\ref{fig:network}.

\begin{figure}[t]
	\centering
	\includegraphics[width=0.95\textwidth]{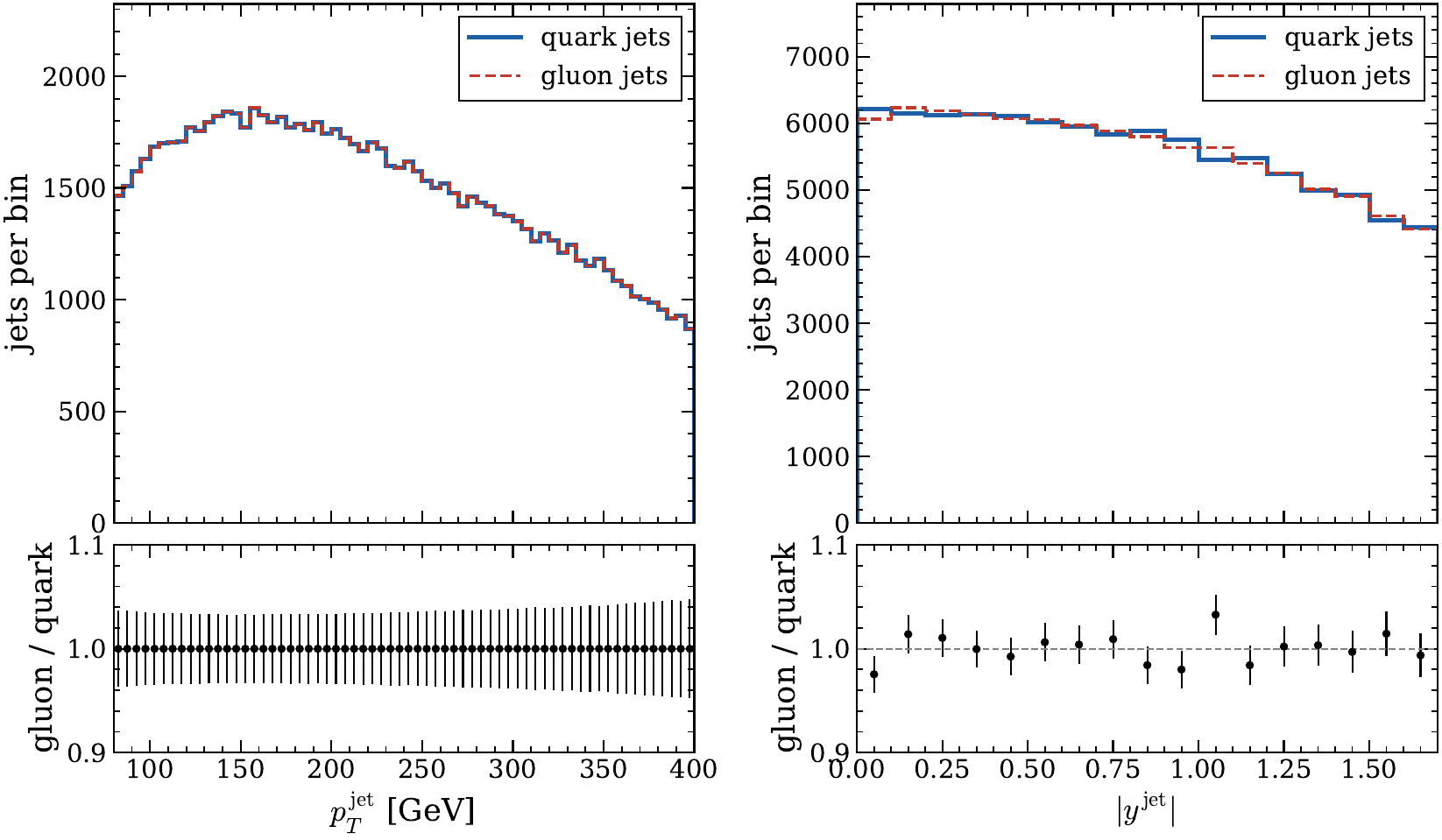}
	\caption{Jet transverse-momentum (left) and rapidity (right) distributions of the quark and gluon samples after the cell-wise matching of eq.~\eqref{eq:target}, with the gluon/quark ratio in the lower panels. The $|y|$ distribution is shown in finer bins than the eight balancing bins.}
	\label{fig:matching}
\end{figure}

The network is trained by minimizing the cross-entropy between this predicted probability and the generator-level truth label. In this way, the trained network effectively learns to exploit differences in the angular and energy-flow patterns of quark- versus gluon-initiated jets, such as the tendency of gluon jets to be broader and to have a higher particle multiplicity due to their larger color charge, without these observables being explicitly engineered as inputs.

\section{Event generation and training}
\label{sec:setup}

\subsection{Event samples}
\label{sec:sample}

The quark and gluon jets used to train and evaluate the network described in section~\ref{sec:gnn} are generated with \textsc{Pythia}~8.318~\cite{Sjostrand:2014zea} for $Z$+jet production at $\sqrt{s}=14$~TeV, following the setup of the standard quark/gluon benchmark of ref.~\cite{Komiske:2018cqr} that was also used for ParticleNet~\cite{Qu:2019gqs}, but now with a substantially wider range of jet transverse momenta. The classes are defined by the hard process,
\begin{equation}
q\bar q\to Z+g\,,\qquad qg\to Z+q\,,
\label{eq:procs}
\end{equation}
with $q\in\{u,d,s\}$. The $Z$ boson decays to neutrinos, such that its decay products do not enter the jet under consideration. All other effects like initial-state radiation (ISR), multi-parton interactions (MPI) and hadronization were turned on for the data generation.

Final-state particles excluding neutrinos are clustered with the anti-$k_t$ algorithm~\cite{Cacciari:2008gp} with $R=0.4$ using \textsc{FastJet}~3.3.0~\cite{Cacciari:2011ma}, and the leading jet is retained if $80\le\ptjet\le400$~GeV and $|y_{\rm jet}|<1.7$, without matching to the hard parton. The hard-scattering transverse momentum is generated with $\hat p_T\ge60$~GeV using a phase-space bias of $4$ in the  \textsc{Pythia} event generation so as to populate the high-$\pT$ regions more effectively. To keep the network size under control, the first 100 highest $\pT$ constituents of the jet are passed to the network.

Furthermore, to avoid that the classifier exploits differences in the production-level kinematics of the two samples, the jets are sub-divided into 64 bins in $\ptjet$ of 5~GeV width and 8 bins in $|y|$, and the two classes are balanced independently in each of the 512 cells. The target occupancy of cell $b$ is then the class average of occupancies $n_{q,b}$ and $n_{g,b}$ of test samples of $10^5$ jets per class, rescaled to the final sample size,
\begin{equation}
T_b=\frac{10^6}{10^5}\,\frac{n_{q,b}+n_{g,b}}{2}\,,
\label{eq:target}
\end{equation}
and the final set of $10^6$ jets per class are subsampled accordingly, that are then used for all the further analysis. 
The two classes then have identical distributions in $(\ptjet,|y|)$ at the level of the cell binning, see fig.~\ref{fig:matching}. 
Each cell then contributes $80\%$ of its jets to the training, $10\%$ to the validation and $10\%$ to the test sample, i.e.\ $1.6\times10^6$, $2\times10^5$ and $2\times10^5$ jets with equal numbers of quark and gluon jets.

It is important to note that while through this matching, we make the marginal distributions of $\ptjet$ and $|y|$ identical for the two classes, such that they carry no discriminating information by themselves, it does not remove their correlation with the substructure. This can be understood as the transverse momentum of the jet, $\ptjet$, can be reconstructed from the constituent momenta. This implies that the network can use the jet scale jointly with the substructure, as is required to adapt the discrimination to the $\pT$ dependence of quark and gluon jets, which we will show later.

\subsection{Network training and output}
\label{sec:training}

The network is trained by minimizing the cross-entropy with AdamW optimizer~\cite{Loshchilov:2017adamw} and a one-cycle learning-rate schedule~\cite{Smith:2017onecycle}, rising from $3\times10^{-4}$ to $3\times10^{-3}$ and decreasing to $5\times10^{-7}$, for 20 epochs with batches of 384 jets. The weights with the best validation accuracy are retained.

For each jet, the network returns two outputs $\ell_q$ and $\ell_g$, and all information is contained in their difference,
\begin{equation}
\Mg\equiv\ell_g-\ell_q\,,\qquad P(g)=\frac{1}{1+e^{-\Mg}}\,.
\label{eq:margin}
\end{equation}
Since the cross-entropy is minimized by the class posterior, and the two classes are equally populated in every kinematic cell, the output of an ideally trained network is the log-likelihood ratio~\cite{Cranmer:2015bka}
\begin{equation}
\Mg^\star(\mathbf{x})=\ln\frac{p(\mathbf{x}\,|\,g)}{p(\mathbf{x}\,|\,q)}\,,
\label{eq:LLR}
\end{equation}
where $\mathbf{x}$ denotes all network inputs. A jet is classified as a gluon jet if $\Mg>0$ and as a quark jet if $\Mg<0$. We further define the true-class output $\Mg_t=\Mg$ for gluon jets and $\Mg_t=-\Mg$ for quark jets, such that a jet is correctly identified if and only if $\Mg_t>0$.

\section{Quark/gluon discrimination with jet observables}
\label{sec:reference}

To quantify what the network learns beyond known jet observables, we first construct the best classifiers that can be built from traditional QCD jet substructure observables. We begin with the constituent multiplicity alone, and then add other standard observables.

\subsection{Multiplicity likelihood ratio}
\label{sec:LN}

\begin{figure}[!tb]
\centering
\includegraphics[width=0.8\textwidth]{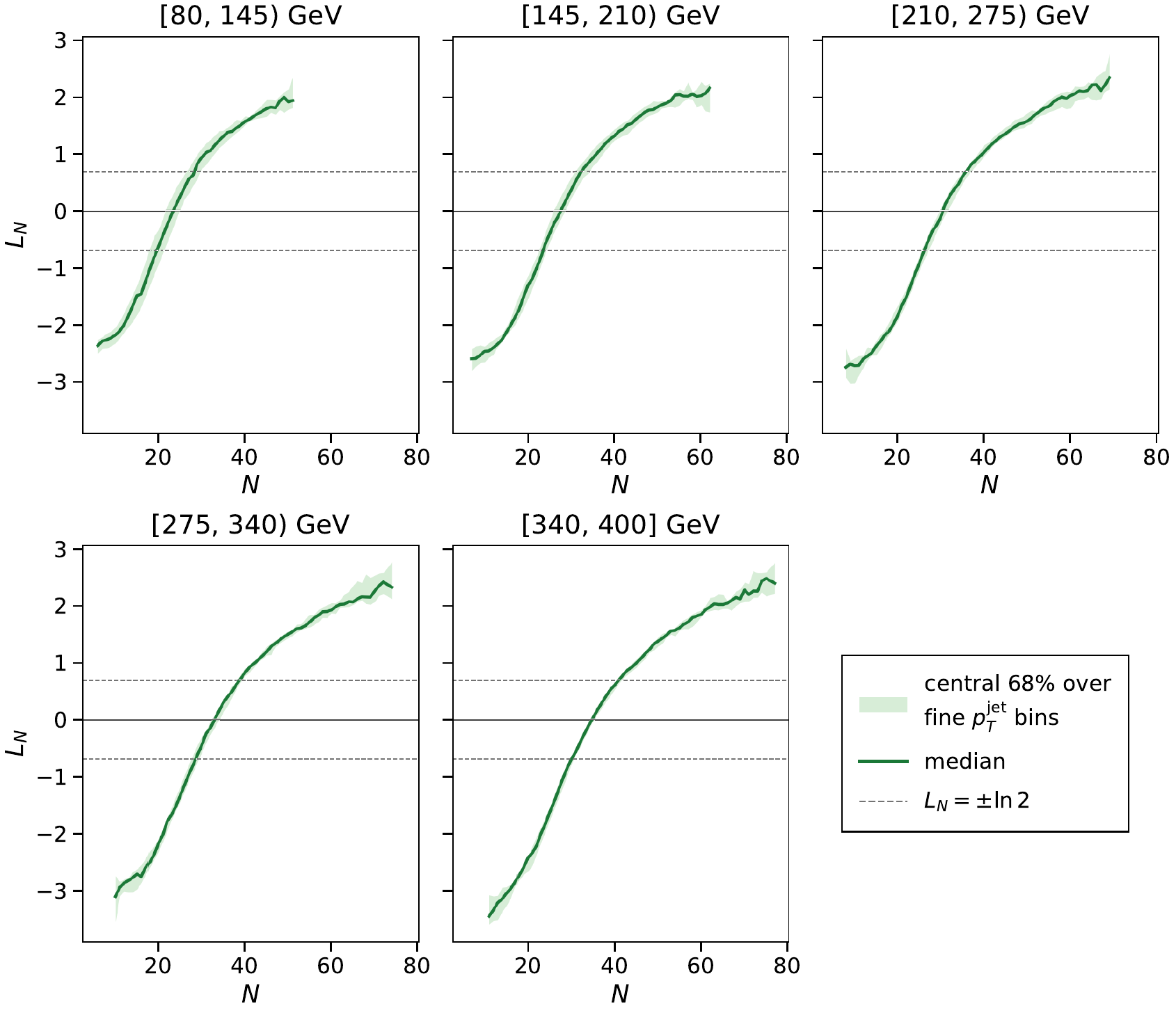}
\caption{Multiplicity-only log-likelihood ratio $\LN$, eq.~\eqref{eq:LN}, estimated from the training sample, as a function of the constituent multiplicity in five $\ptjet$ bins. The line shows the median over the fine $\pT$ bins populated at each $N$ and the band the central 68\%; the dashed lines indicate $\LN=\pm\ln2$. }
\label{fig:LN}
\end{figure}

The optimal classifier based on multiplicity alone is the log-likelihood ratio~\cite{Neyman:1933wgr,Larkoski:2019nwj}
\begin{equation}
\LN(N,\ptjet)=\ln\frac{p(N\,|\,g,\ptjet)}{p(N\,|\,q,\ptjet)}\,,
\label{eq:LN}
\end{equation}
which we estimate from the training sample in 64 bins of $\ptjet$ of 5~GeV width, with a mild smoothing in $N$ to stabilize the tails of the distributions ( see app.~\ref{app:LN} for details). Neither the network output nor its decisions enter this construction. From eq.~\eqref{eq:LN}, we note that $\LN>0$ when the observed multiplicity is more likely for a gluon jet and $\LN<0$ if it is more probably a quark jet. Figure~\ref{fig:LN} shows $\LN$ as a function of $N$ in five bins of $\ptjet$. From the figure, we observe that $\LN$ increases monotonically with $N$, its zero moves to higher multiplicity with increasing $\ptjet$, and its slope increases.

To quantify the discriminating power of the multiplicity, we use the area under the relative operating characteristic (ROC) curve. This area under the ROC curve which we refer as the AUC equals the probability that a randomly chosen gluon jet is assigned a more gluon-like value of the classifier than a randomly chosen quark jet~\cite{Hanley:1982auc}. ${\rm AUC}=0.5$, thus, corresponds to no separation and ${\rm AUC}=1$ to a perfect one. 

\begin{table}[h]
	\centering
	\begin{tabular}{|p{2cm}|p{2cm}|p{2cm}|p{2cm}|p{2cm}|p{2cm}|}
		\hline
		$\ptjet$ [GeV] & 80--145 & 145--210 & 210--275 & 275--340 & 340--400\\
		\hline
		$N$ & 0.783 & 0.808 & 0.817 & 0.830 & 0.841\\
		\hline
		$\LN$ & 0.787 & 0.809 & 0.818 & 0.830 & 0.841\\
		\hline
	\end{tabular}
	\caption{AUC of multiplicity and the multiplicity likelihood ratio in five bins of $\ptjet$, evaluated on the validation sample.}
	\label{tab:auc}
\end{table}

Table~\ref{tab:auc} shows the AUC of $N$ and $\LN$ in five bins of $\ptjet$ for the jets of the validation sample. We see that multiplicity alone separates quark and gluon jets with an AUC between $0.783$ in the lowest and $0.841$ in the highest $\ptjet$ bin. The values for $N$ and $\LN$ are very close, since both use the same information. However, $\LN$ is marginally higher in the lowest bins because, unlike $N$, it also knows the jet transverse momentum, and can therefore account for the shift of multiplicity distributions towards larger $N$ with increasing $\ptjet$ within a bin. This effect is largest in the lowest bin, where $\ptjet$ varies by almost a factor of two, and negligible in the highest one. 

Both classifiers improve with increasing $\ptjet$, as the difference between mean multiplicities of gluon and quark jets grows with the phase space available for radiation faster than their spread~\cite{Frye:2017yrw,Larkoski:2014pca}. In the following, we, therefore, use $\LN$ as the reference for the discriminating power of multiplicity alone, against which  global jet observables and the network are compared.

\subsection{Global jet observables and a boosted-decision-tree baseline}
\label{sec:bdt}

\begin{figure}[t]
	\centering
	\includegraphics[width=0.9\textwidth]{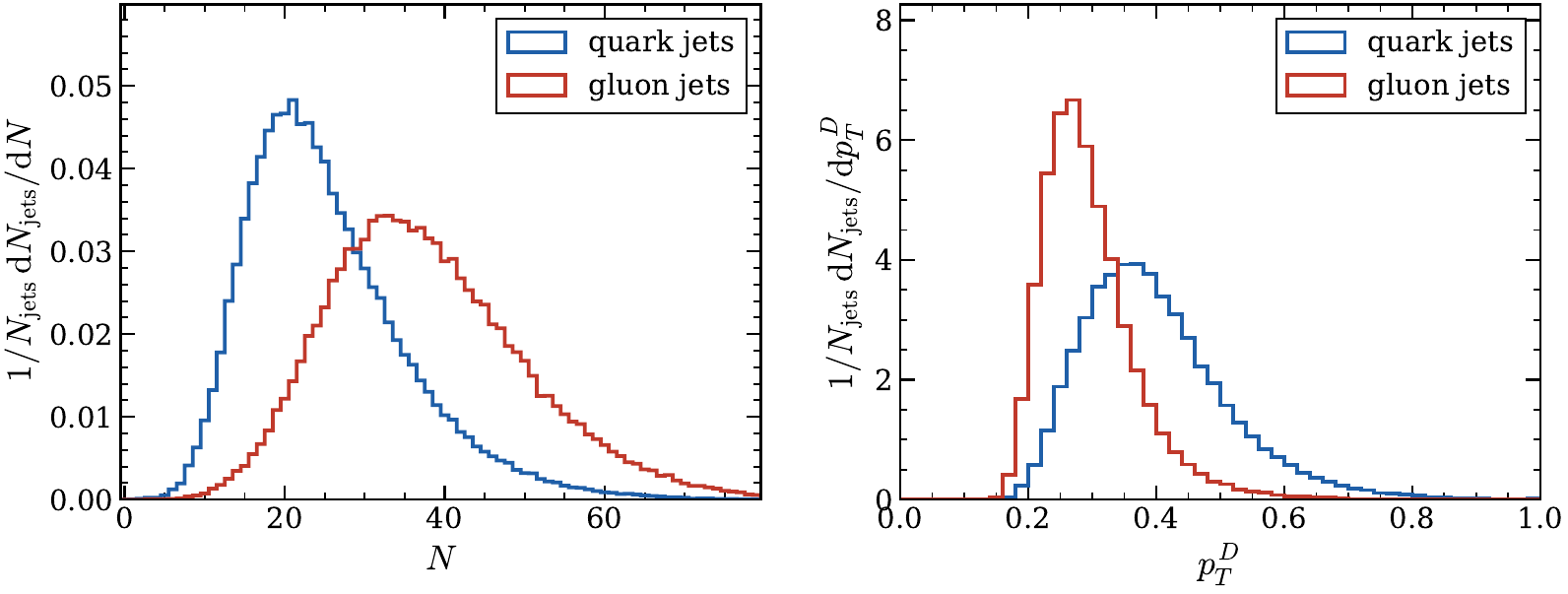}\\[2mm]
	\includegraphics[width=0.9\textwidth]{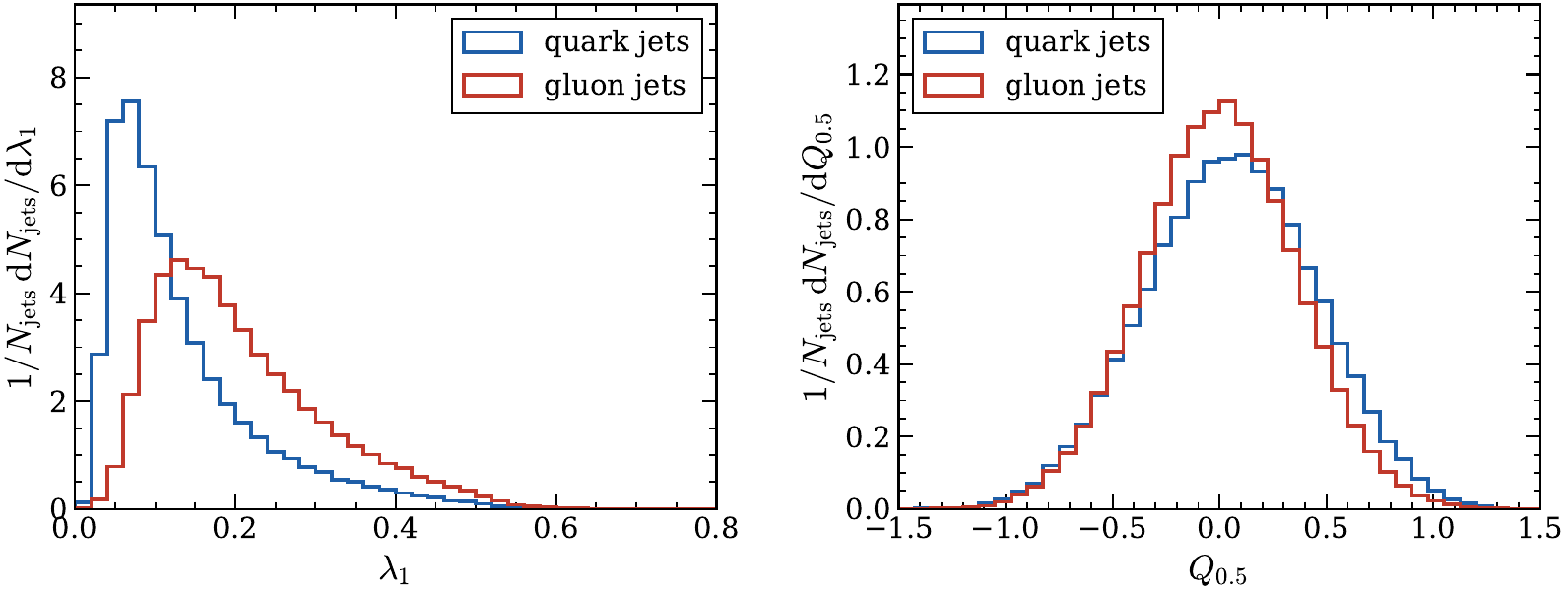}
	\caption{Normalized distributions of constituent multiplicity $N$, momentum dispersion $p_T^D$ (top), and for jet angularity $\lambda_1$ and  jet charge $Q_{0.5}$ (bottom), for quark and gluon jets from the \textsc{Pythia} samples after  matching in $(\ptjet,|y|)$, for $80\le\ptjet<400$~GeV..}
	\label{fig:dists}
\end{figure}

Besides larger multiplicities, gluon jets have a radiation pattern distinct from that of quark-initiated jets, for instance, they tend to be broader and have their momentum shared amongst more soft components. To this end, we consider three jet substructure observables sensitive to these differences, the momentum dispersion $p_T^D$~\cite{CMS:2013kfa}, the angularities $\lambda_\beta$~\cite{Berger:2003iw,Ellis:2010rwa,Larkoski:2014pca} and the jet charge $Q_\kappa$~\cite{Field:1977fa,Waalewijn:2012sv,Krohn:2012fg},
\begin{equation}
p_T^D=\frac{\sqrt{\sum_i z_i^2}}{\sum_i z_i}\,,\qquad
\lambda_\beta=\sum_i z_i\left(\frac{r_i}{R}\right)^{\beta}\,,\qquad
Q_\kappa=\sum_i q_i\, z_i^{\kappa}\,,
\label{eq:obs}
\end{equation}
where the sums run over the jet constituents, $z_i=p_{T,i}/\ptjet$ is the momentum fraction of constituent $i$, $r_i$ its distance to the jet axis in the rapidity--azimuth plane and $q_i$ its electric charge. While $p_T^D$ is large when a few constituents carry most of the jet momentum, as is more typical of quark jets, angularities measure how the momentum is distributed in angle: $\lambda_1$ is the jet broadening, $\lambda_2$ is closely related to the jet mass.\footnote{For $\beta>0$ the angularities are infrared and collinear safe, whereas $N$ and $p_T^D$ are not and depend on the hadronization.} Similarly, the average jet charge is zero for gluon jets and has the sign of the charge of the initiating quark for quark jets. Since our sample contains quarks and antiquarks of different flavors, the discrimination with $Q_\kappa$ is expected to be weak. The normalized distributions of these jet observables are shown in fig.~\ref{fig:dists} for quark jets (blue) and gluon jets (red).

To obtain a baseline that combines these observables, we train a BDT on $N$, $\ptjet$, $p_T^D$, $\lambda_1$, $\lambda_{0.5}$ and $Q_{0.5}$, computed from the same constituents as the network input. The BDT is trained with \textsc{XGBoost}~3.0.5~\cite{Chen:2016xgb} on the training sample of section~\ref{sec:sample}, using a binary logistic objective, histogram training with 256 bins, maximum tree depth 4, learning rate 0.05, row subsampling 0.8, minimum child weight 20 and L2 leaf regularization 5. The number of trees (2341) is chosen by maximizing the AUC on a hold-out subset of $1.6\times10^5$ training jets, after which the BDT is refitted on the full training sample. Like the network, the BDT is trained on all transverse momenta at once, and the inclusion of $\ptjet$ allows it to adapt to the $\pT$ dependence of the observables.

\begin{table}[h]
	\centering
	\begin{tabular}{|p{4.2cm}|p{1.3cm}|p{1.5cm}|p{1.5cm}|p{1.5cm}|p{1.5cm}|}
		\hline
		$\ptjet$ [GeV] & 80--145 & 145--210 & 210--275 & 275--340 & 340--400\\
		\hline
		$\LN$ & 0.787 & 0.809 & 0.818 & 0.830 & 0.841\\
		\hline
		BDT classifier trained on ($N,\ptjet,p_T^D,\lambda_1,\lambda_{0.5},Q_{0.5}$) & 0.807 & 0.829 & 0.837 & 0.853 & 0.865\\
		\hline
	\end{tabular}
	\caption{AUC of the boosted decision tree built from six jet observables, compared to the multiplicity likelihood ratio, in five bins of $\ptjet$.}
	\label{tab:bdt}
\end{table}

Table~\ref{tab:bdt} compares the AUC of the BDT with that of the multiplicity likelihood ratio. The BDT improves on $\LN$ in every $\ptjet$ bin. 
This improvement is small compared to the separation provided by multiplicity itself, and the BDT, like $\LN$, improves steadily with increasing $\ptjet$. We will show in section~\ref{sec:auc} that the improvement of the network over $\LN$ is about three times larger, such that these six observables roughly account for about one third of what the network learns beyond multiplicity.

\subsection{Correlations among jet observables}
\label{sec:indep}

A single BDT classifier does not reveal how the quark/gluon discrimination ability is distributed among the observables. Observables such as $N$, $p_T^D$ and the angularities all measure, to a large extent, how much the initiating parton has radiated, and are therefore strongly correlated. A small gain from adding an observable to the multiplicity can thus mean either that the observable is weakly discriminating, or that most of its information is already contained in $N$. To distinguish the two, we compare classifiers built from different sets of observables. In addition to $N$, we consider $p_T^D$, $\lambda_{0.5}$, $\lambda_1$, $\lambda_2$, $m/\ptjet$ and $Q_{0.5}$. For each set of observables, we train a BDT~\cite{Pedregosa:2011sk}, always including $\ptjet$ and $|y|$, and evaluate each jet with a BDT trained on the remaining jets of the same $\ptjet$ bin.

\begin{table}[h]
\centering
\begin{tabular}{|p{1.6cm}|p{1.0cm}|p{1.0cm}|p{1.0cm}|p{1.2cm}|p{1.2cm}|p{1.6cm}|p{1.3cm}|p{1.4cm}|}
\hline
$\ptjet$ [GeV] & $N$ & $p_T^D$ & $\lambda_1$ & $N,\lambda_1$ & $N,p_T^D$ & $N,\lambda_1,p_T^D$ & all seven & all but $N$\\
\hline
80--145  & 0.784 & 0.773 & 0.741 & 0.796 & 0.793 & 0.802 & 0.806 & 0.800\\
\hline
210--275 & 0.818 & 0.795 & 0.744 & 0.826 & 0.822 & 0.833 & 0.838 & 0.827\\
\hline
340--400 & 0.842 & 0.812 & 0.749 & 0.851 & 0.844 & 0.857 & 0.862 & 0.846\\
\hline
\end{tabular}
\caption{AUC of single jet observables and of BDT classifiers built from sets of jet observables ($\ptjet$ and $|y|$ always included) in three bins of $\ptjet$.}
\label{tab:ladder}
\end{table}

\begin{figure}[t]
\centering
\includegraphics[width=0.55\textwidth]{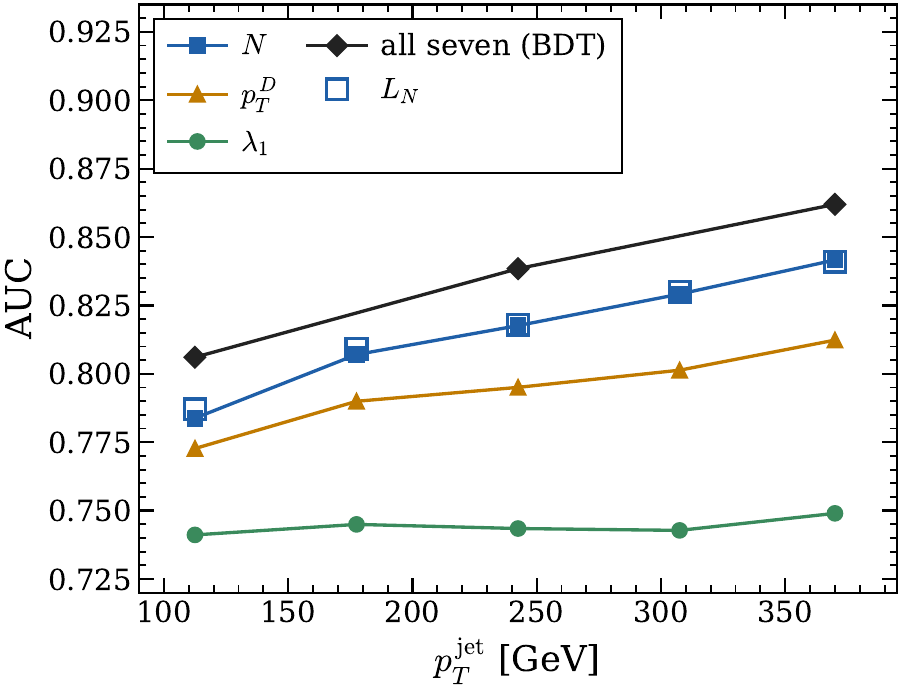}
\caption{AUC of single jet observables and of the BDT classifier on all seven observables as a function of $\ptjet$ compared to the multiplicity likelihood ratio $\LN$ of table~\ref{tab:auc}.}
\label{fig:ladder}
\end{figure}

The results are tabulated in Table~\ref{tab:ladder} and shown in fig~\ref{fig:ladder}. We find that the momentum dispersion alone is nearly as discriminating as multiplicity at low $\ptjet$, while angularities reach $0.74$--$0.75$ and jet charge only $0.53$--$0.55$. Nevertheless, adding $\lambda_1$ and $p_T^D$ to multiplicity increases the AUC by only $0.015$--$0.018$, and all seven observables together gain $\sim 0.02$ over $N$, comparable to the gain of BDT over $\LN$ in table~\ref{tab:bdt}. Furthermore, we find that, $\lambda_1$ adds more to multiplicity than $p_T^D$, although $p_T^D$ is the stronger discriminant on its own. This is because $\lambda_1$ provides more distinct information of the jet structure than $p_T^D$ when added to multiplicity. 

Conversely, removing $N$ from the full set lowers the AUC by only $0.006$ at $80$--$145$~GeV but by about $0.016$ at $340$--$400$~GeV which is almost the same size as the gain obtained by adding observables. This suggests that although most of the information carried by $N$ is shared with $p_T^D$ and angularities,  at high $\ptjet$ the multiplicity alone carries more information of its own.

\section{Quark vs.\ gluon performance of the network}
\label{sec:perf}

We now analyze the discrimination provided by the network and subsequently first use our trained model in the evaluation mode and investigate what features about the jet structure the network classifies indirectly by comparing against our expected distributions. Finally, we turn to providing a probabilistic interpretation of the network's output by directly analyzing the logit response of the network. 

\subsection{Discrimination power}
\label{sec:auc}

\begin{figure}[!htb]
    \centering
    \includegraphics[width=0.9\linewidth]{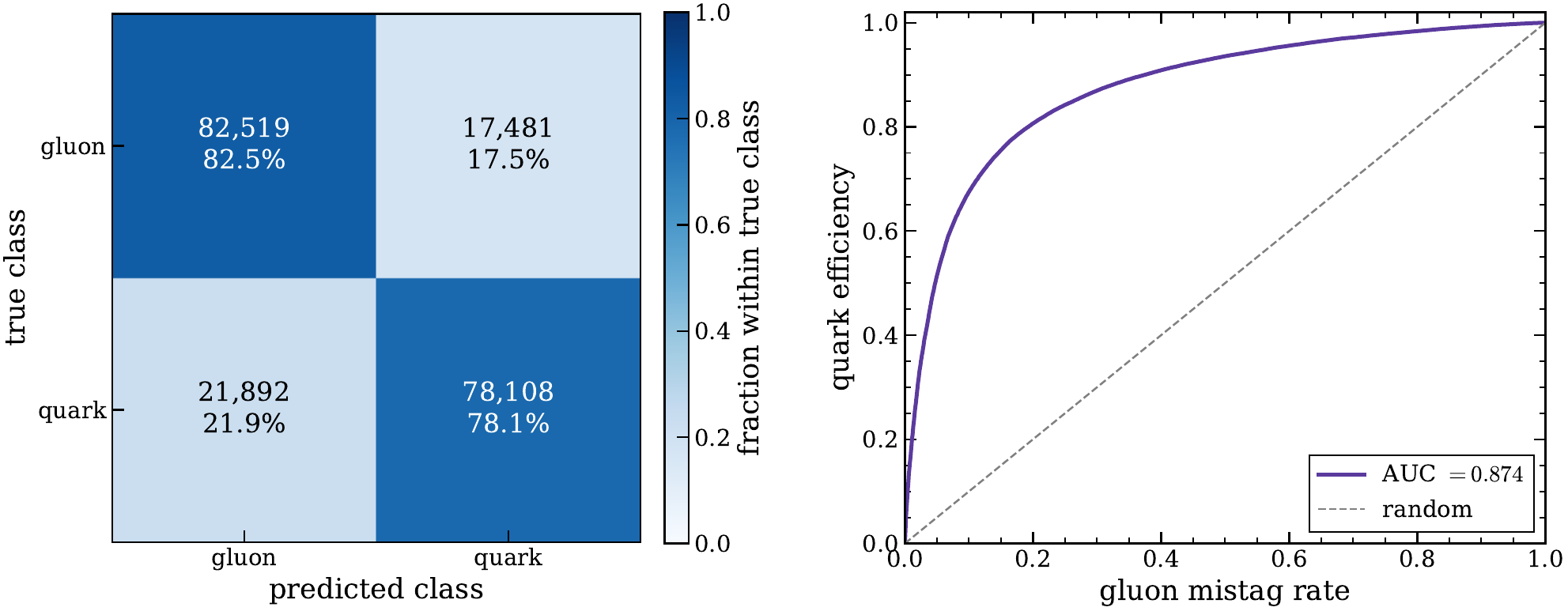}
    \caption{Confusion matrix at $P(q)\ge0.5$ (left) and ROC curve (right) of the network on the test sample.}
    \label{fig:Overall-ROC}
\end{figure}

We first quantify the quark-gluon discrimination learned by our graph neural network over the full jet transverse momentum range and subsequently study how the learned discrimination changes when the jets are classified more differentially, i.e., on the basis of their transverse momentum and constituent multiplicity. Fig.~\ref{fig:Overall-ROC} shows the confusion matrix at a fixed threshold of $P(q) > 0.5$ and the ROC curve evaluated on the test set of $2\times10^5$ jets that were not used during the training process. The overall accuracy achieved by the network over this test set is 0.803, with 78.1\% of quark jets and 82.5\% of gluon jets correctly identified, and the AUC is 0.874.

\begin{figure}[!htb]
	\centering
	\includegraphics[width=0.8\linewidth]{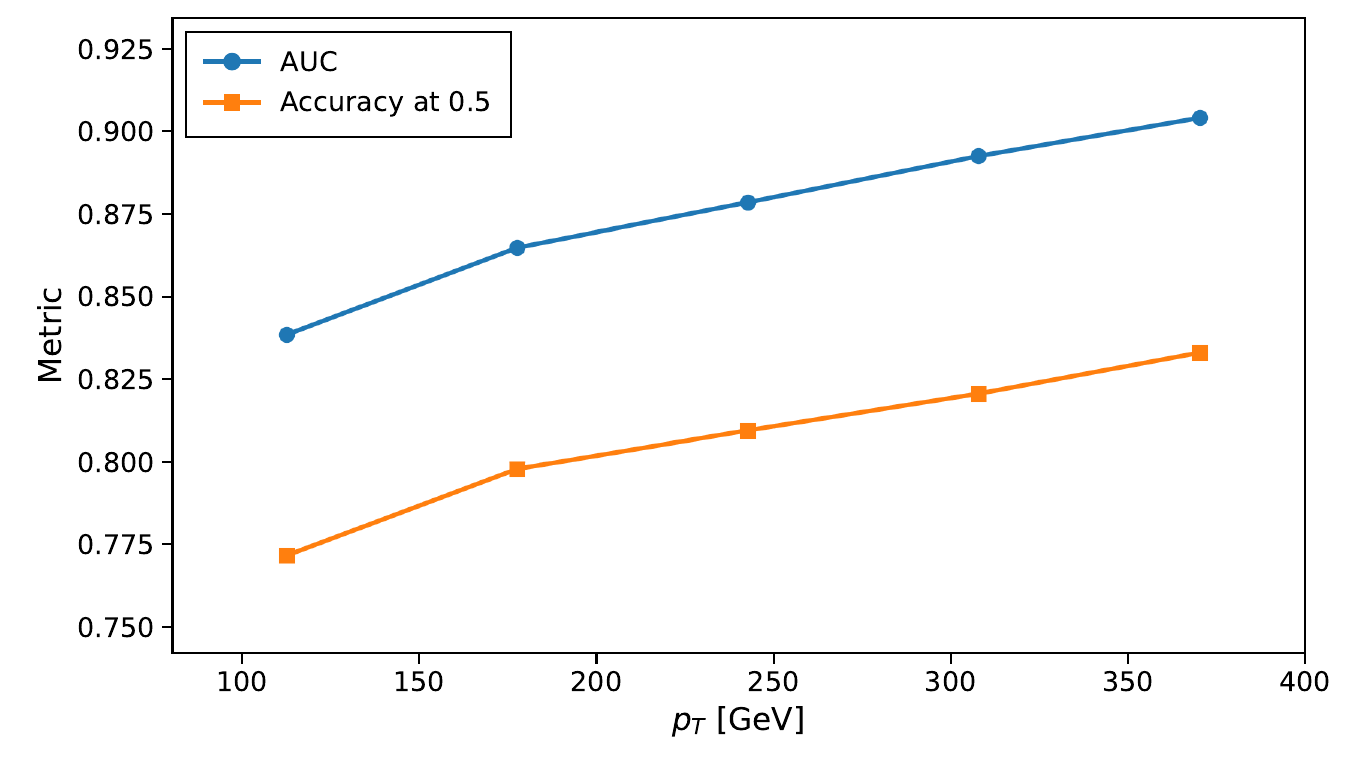}
	\caption{AUC and accuracy at $P(g)=0.5$ of the network in five bins of $\ptjet$ evaluated on the test sample.}
	\label{fig:AUC-PT}
\end{figure}

Next, we evaluate the network performance in our five bins of $\ptjet$. The AUC along with network accuracy at a fixed threshold of $0.5$ is shown in fig~\ref{fig:AUC-PT}. We find that the AUC increases from $0.84$ in the lowest to $0.90$ in the highest bin. The multiplicity likelihood ratio shows the same increase, from $0.787$ to $0.841$, and the gain of the network over multiplicity is nearly constant: on the validation sample, on which table~\ref{tab:auc} is evaluated, the network reaches an AUC of $0.842$, $0.865$, $0.878$, $0.894$ and $0.904$, i.e.\ $0.055$--$0.064$ above $\LN$. The observed increase with $\ptjet$ is similar to what we expected as explained in sec.~\ref{sec:LN}~\cite{Frye:2017yrw,Larkoski:2014pca}. Moreover, the inclusive AUC is an average over the $\ptjet$ bins and hence is reduced due to the inclusion of low $p_T$ bins where the overlap of quark vs gluon class is itself large.

\begin{table}[t]
	\centering
	\begin{tabular}{|p{4.2cm}|p{1.5cm}|p{1.5cm}|p{1.5cm}|p{1.5cm}|p{1.5cm}|}
		\hline
		$\ptjet$ [GeV] & 80--145 & 145--210 & 210--275 & 275--340 & 340--400\\
		\hline
		$\LN$ & 0.787 & 0.809 & 0.818 & 0.830 & 0.841\\
		\hline
		BDT on six observables & 0.807 & 0.829 & 0.837 & 0.853 & 0.865\\
		\hline
		\textsc{ParticleNet} & 0.842 & 0.865 & 0.878 & 0.894 & 0.904\\
		\hline
		\hline
		gain of BDT over $\LN$ & 2.5\% & 2.4\% & 2.4\% & 2.8\% & 2.8\%\\
		\hline
		gain of \textsc{ParticleNet} over $\LN$ & 7.0\% & 6.8\% & 7.4\% & 7.7\% & 7.5\%\\
		\hline
	\end{tabular}
	\caption{AUC of the multiplicity likelihood ratio, of the BDT and of \textsc{ParticleNet} in five bins of $\ptjet$ (validation sample), together with the relative improvement of the two classifiers over $\LN$. The gain achieved by \textsc{ParticleNet} over multiplicity alone is about three times larger than that of the BDT.}
	\label{tab:gain}
\end{table}

Finally, in table~\ref{tab:gain} we quantify the gain attained by \textsc{ParticleNet} over multiplicity alone, and compare it to the gain of BDT built from six standard jet observables of sec~\ref{sec:bdt}. In every $\ptjet$ bin, \textsc{ParticleNet} improves the AUC of the multiplicity likelihood ratio by $7$--$8\%$, whereas the BDT improves it by only $2$--$3\%$. The gain of \textsc{ParticleNet} is thus about three times larger, and remains nearly constant across our $\ptjet$ range. This shows that the network exploits substantial information in the jet beyond multiplicity. We investigate this information in sec~\ref{sec:beyond}.

\begin{figure}[t]
	\centering
	\includegraphics[width=0.9\linewidth]{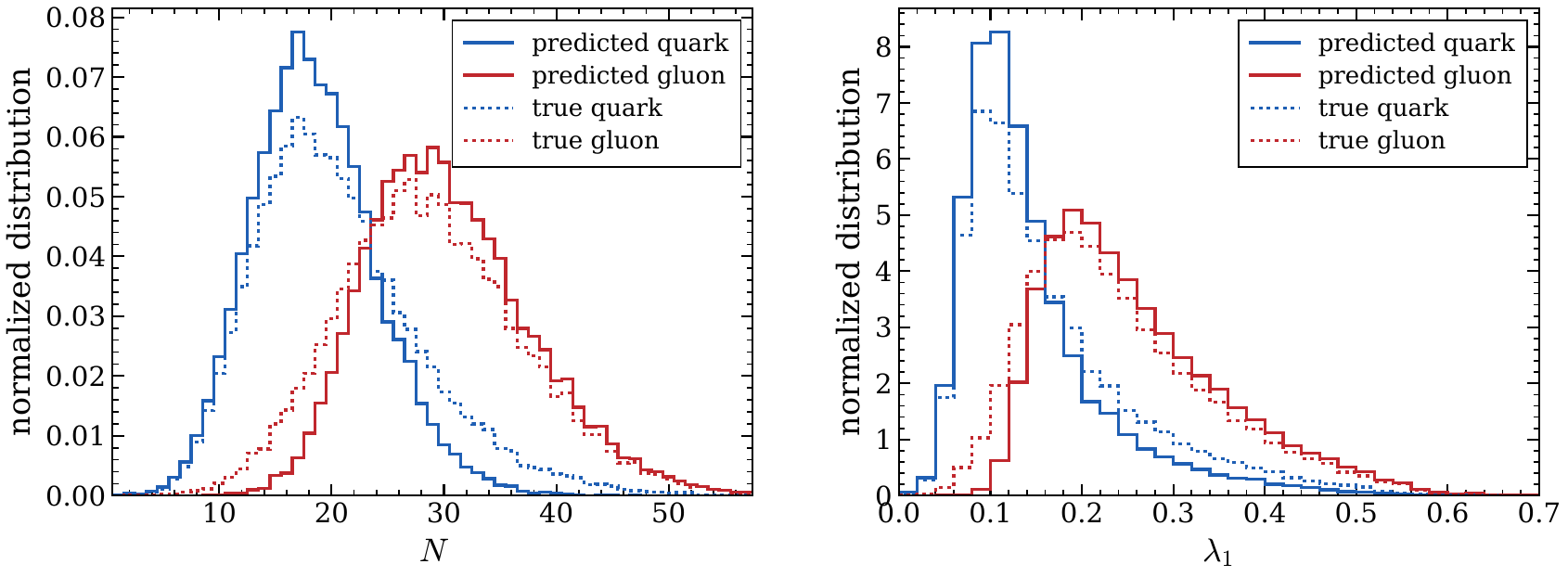}
	\caption{Distributions of constituent multiplicity $N$ and broadening $\lambda_1$ for true (dotted) and predicted (solid) quark and gluon jets in the bin $80\le\ptjet\le145$~GeV evaluated on the validation sample.}
	\label{fig:obspred}
\end{figure}

\subsection{Substructure of quark- vs gluon-jets as learnt by the network}
\label{sec:pred}

Before interpreting the network output itself, we first look at the jet discrimination learnt by the network. For this, we use our trained network in evaluation mode on the validation sample and assign each jet to the network's predicted quark or gluon class. We then compare the distributions of jet substructure observables in the predicted classes against the truth labels.

The resulting distributions are shown for $N$ and $\lambda_1$ in fig.~\ref{fig:obspred} for the true and the predicted classes. We find that the predicted distributions closely resemble the truth ones. However, the predicted classes are somewhat more strongly separated as visible more clearly in the multiplicity and to some extent also in $\lambda_1$ distributions. This would be explained if the network discrimination strongly relies on the multiplicity of jets, i.e. a quark jet with high multiplicity (typical of gluon jets), or a gluon jet with a multiplicity typical of a quark jets would then be assigned to the opposite class. As a result, the tails of the truth distributions would get moved to the opposite class.

\begin{figure}[!htb]
    \centering
    \includegraphics[width=1\linewidth]{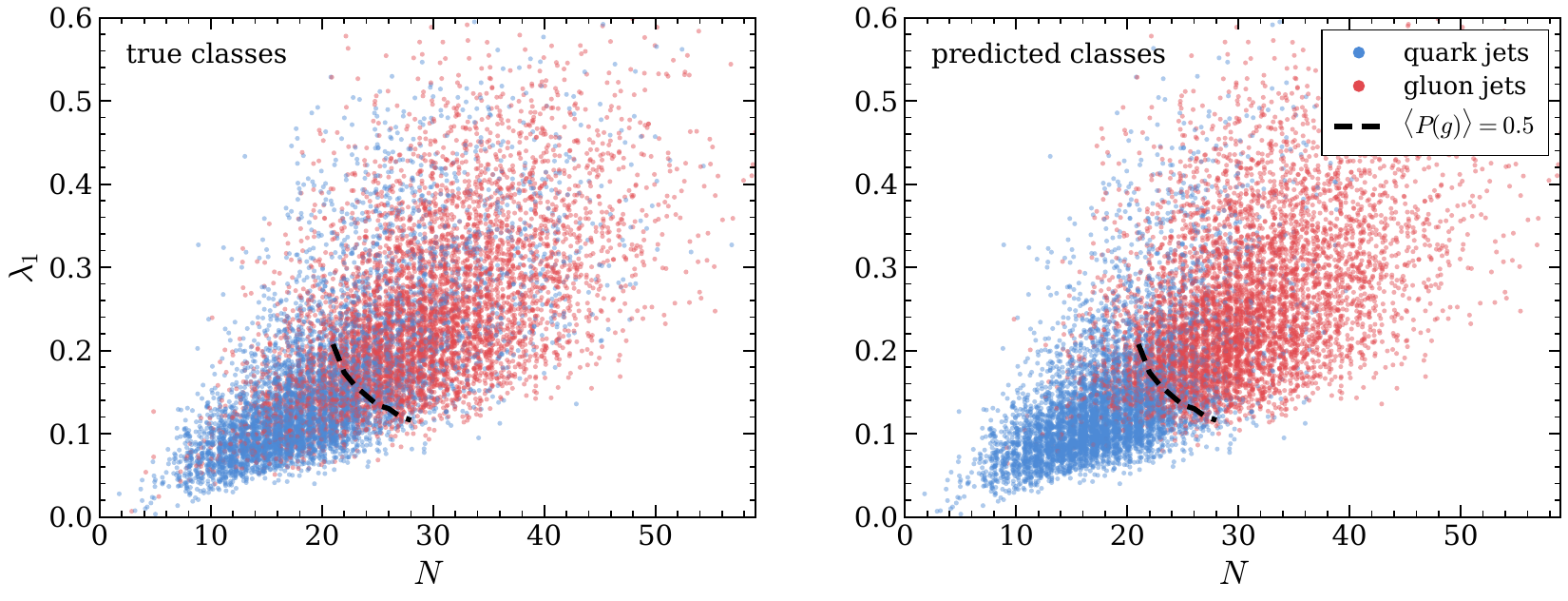}
    \caption{Scatter plots of constituent multiplicity $N$ and the broadening $\lambda_1$ for the truth (left) and predicted (right) classes in the bin $80\le\ptjet\le145$~GeV evaluated on the validation sample.}
    \label{fig:Nl1}
\end{figure}

Next, in fig~\ref{fig:Nl1}, we also look at the correlations between these observables. To this end, we only show the correlations between constituent multiplicity $N$ and angularity $\lambda_1$ since among the observables of section~\ref{sec:indep} $\lambda_1$ adds the most discriminating power to $N$. For the truth labels, fig~\ref{fig:Nl1} (left), $N$ and $\lambda_1$ are positively correlated, since both grow with the amount of radiation in the jet. We find that this correlation is also present in the predicted classes, see fig~\ref{fig:Nl1} (right). However, the network assigns regions of large $N$ and large $\lambda_1$ mostly to gluons and the region of small $N$ and small $\lambda_1$ as quarks. 
We will focus on these class of misidentified jets to explore what features beyond multiplicity allow the network to correctly identify these class of jets, in sec.~\ref{sec:beyond}.

Before that, let us consider what extent of the classification of the network is governed by multiplicity alone directly through the use of network outputs in the section below.

\subsection{Probabilistic interpretation of the network output}
\label{sec:prob}

\begin{figure}[t]
	\centering
	\includegraphics[width=\textwidth]{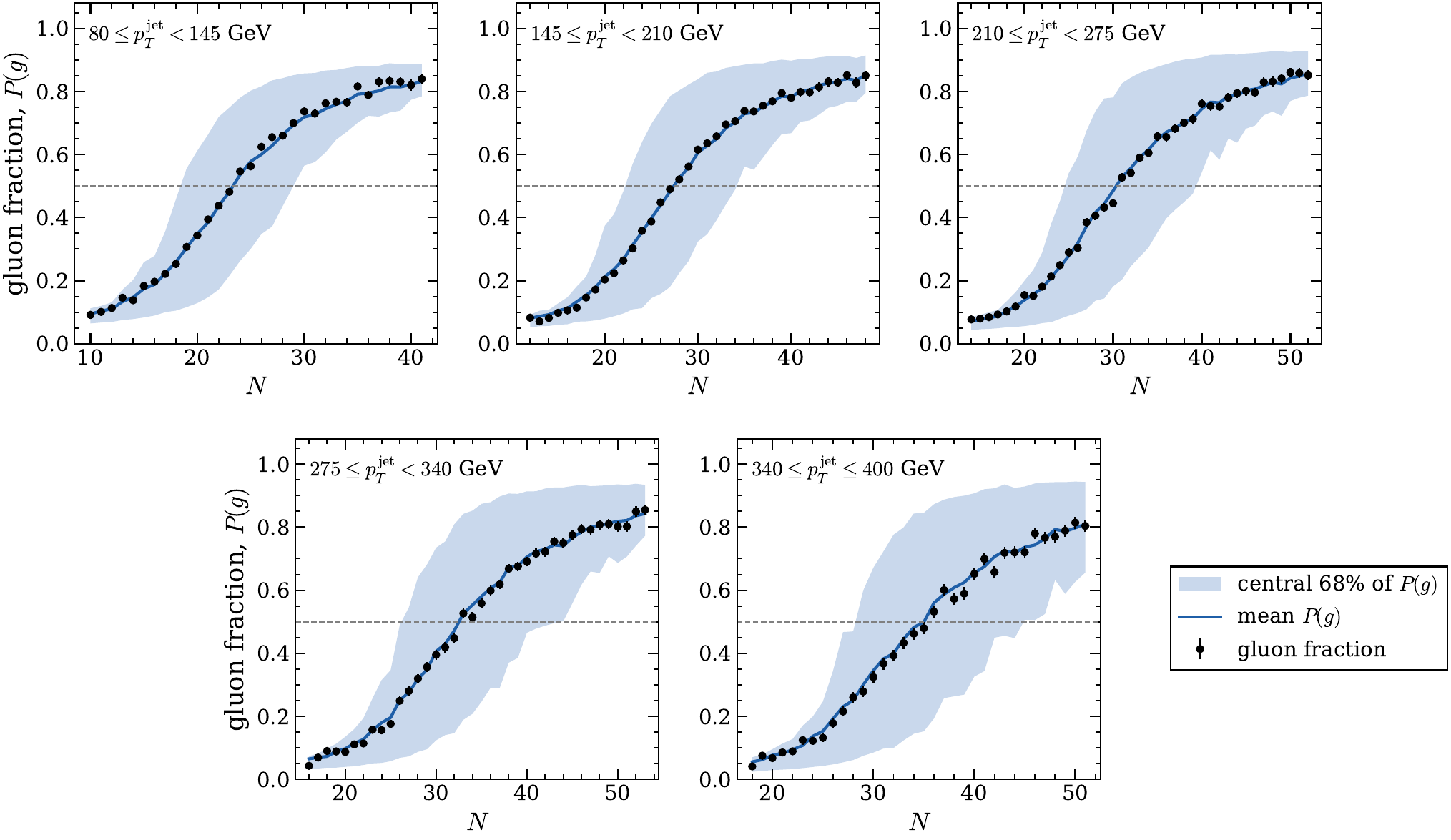}
	\caption{Fraction of gluon jets (dots) and mean gluon probability $P(g)$ assigned by the network (solid curve) as a function of the constituent multiplicity in five bins of $\ptjet$. The band shows the central 68\% of $P(g)$ at each $N$, and the dashed line marks $P(g)=0.5$.}
	\label{fig:pgN}
\end{figure}

For each jet, the network returns two outputs, $\ell_q$ and $\ell_g$, and classifies the jet as a gluon jet if their difference $M=\ell_g-\ell_q$ is positive, i.e. if
\begin{equation}
P(g)=1/(1+e^{-M})>0.5 \, .
\end{equation}
For an ideally trained network, the output $M$ is the log-likelihood ratio in eq.~\eqref{eq:LLR}, and the gluon probability $P(g)$ is then the probability that a jet with the observed constituents was initiated by a gluon. This means that among all jets assigned by the network with $P(g)=0.7$, about $70\%$ should be gluon jets, irrespective of how these jets were selected. This is a much stronger requirement than having a large AUC, which only tells whether gluon jets receive larger values of $M$ than quark jets~\cite{Guo:2017calibration}. If the above requirement holds, then the value of $M$ itself acquires a physical meaning, i.e. $\exp({M})$ becomes the ratio of the gluon to quark probability of the jet. Hence, a jet with $M=\ln2$, for instance, is twice as likely to be a gluon jet than a quark jet. As the multiplicity carries most of the discriminating power, see sec~\ref{sec:LN}, we test this requirement for jets with fixed multiplicities.

If $P(g)$ is the probability, its average over all jets of a given multiplicity must equal the fraction of gluon jets at that multiplicity. Since $N$ is fixed by the network inputs $\mathbf{x}$, this gives
\begin{equation}
\big\langle P(g\,|\,\mathbf{x})\big\rangle_N=P(g\,|\,N)=\frac{1}{1+e^{-\LN}}\,.
\label{eq:avgN}
\end{equation}
This result holds independently of what further information the network relies on. In fig.~\ref{fig:pgN}, we test this by comparing the fraction of true gluon jets (dots) with the mean gluon probability assigned by the network to the same jets (solid curve) for a given $N$, in all five $\ptjet$ bins.

First, we find that eq.~\eqref{eq:avgN} holds reasonably well for all $\ptjet$ bins and the multiplicity at which quark and gluon jets are equally likely (i.e. the dashed horizontal line in the plot) moves from $N\simeq23$ in the lowest to $N\simeq35$ in the highest $\ptjet$ bin. This shows that, on average, the probabilities assigned by the network agree with the observed gluon fractions. The band in fig.~\ref{fig:pgN} instead represents the spread of the probabilities assigned by the network. Since the network performs better at quark/gluon jet discrimination than the multiplicity alone (see sec.~\ref{sec:auc}), this spread reflects physical differences between jets and not merely fluctuations of the network output.
We study these residual features beyond multiplicity in the next section.

\section{Discrimination beyond multiplicity}
\label{sec:beyond}

In the previous section, we found that the network has learnt how the fraction of gluon jets changes with the multiplicity, and that jets of the same multiplicity nevertheless receive a broad range of gluon probabilities. In this section, we separate these two pieces of the network output, and then ask to which physical property of the jet the second piece responds.

\subsection{Multiplicity and residual log-likelihood ratio}
\label{sec:decomp}

Since the multiplicity $N$ and the jet transverse momentum $\ptjet$ are both determined by the constituents of the jet, the probability to observe a jet with inputs $\mathbf{x}$ in class $c$ can be written using the Bayes' theorem as
\begin{equation}
p(\mathbf{x}\,|\,c)=p(\ptjet\,|\,c)\;p(N\,|\,\ptjet,c)\;p(\mathbf{x}\,|\,N,\ptjet,c)\,.
\label{eq:factor}
\end{equation}
The first factor is the same for quark and gluon jets due to the kinematic matching of section~\ref{sec:sample}, and drops out of the ratio in eq.~\eqref{eq:LLR}. The second factor gives the multiplicity likelihood ratio $\LN$ of eq.~\eqref{eq:LN}. The output of an ideally trained network therefore splits into two terms,
\begin{equation}
M^\star(\mathbf{x})=\LN(N,\ptjet)+\Delta^\star(\mathbf{x})\,,\qquad
\Delta^\star(\mathbf{x})=\ln\frac{p(\mathbf{x}\,|\,g,N,\ptjet)}{p(\mathbf{x}\,|\,q,N,\ptjet)}\,.
\label{eq:decomp}
\end{equation}
The first term tells how much more likely the observed number of constituents is for a gluon jet than for a quark jet, and the second how much more likely the remaining structure of the jet is, i.e. how the constituents are arranged in angle, share the jet momentum and carry charge and particle type, once the multiplicity is known. 

For our network, we define the residual log-likelihood ratio as
\begin{equation}
\Delta\equiv M-\LN\,,
\label{eq:Delta}
\end{equation}
which reduces to $\Delta^\star$ to the extent that $M$ has the probabilistic interpretation of the log-likelihood ratio, as discussed in sec~\ref{sec:prob}. If the network relied on multiplicity alone, it would give $\Delta=0$ for all jets, and all jets of same multiplicity would receive the same gluon probability. However, the spread of gluon probabilities at fixed $N$ in fig.~\ref{fig:pgN} already suggests that $\Delta$ is nonzero and differs between jets of the same multiplicity.\footnote{Within one of our $\ptjet$ bins, $\LN$ at fixed $N$ still varies with the fine $\ptjet$ bin, but with a standard deviation of at most $0.28$, compared to $0.5$--$1.6$ for $\Delta$.}

Eq.~\eqref{eq:decomp} also makes explicit when a jet whose multiplicity is typical of the opposite class can be correctly identified. A jet is classified as a gluon jet if $M>0$, i.e. if $\Delta>-\LN$. A quark jet with a gluon-like multiplicity, $\LN>0$, is therefore correctly identified only if $\Delta<-\LN$, i.e. if the remaining structure of the jet is more quark-like than its multiplicity is gluon-like, and correspondingly for gluon jets with a quark-like multiplicity. These are precisely the jets of fig.~\ref{fig:Nl1} that the network assigns to the opposite class. It is important to note that this holds for the ideal classifier as well: whenever $\Delta^\star$ does not overcome $\LN$, such a jet is more likely to belong to the other class, and its misidentification reflects the overlap of quark and gluon jets rather than a deficiency of the network.

To quantify how often the residual term overcomes the multiplicity, we consider the jets whose multiplicity favors the opposite class by odds of at least $2{:}1$, i.e.\ gluon jets with $\LN<-\ln2$ and quark jets with $\LN>\ln2$. A classifier based on the multiplicity alone misidentifies all of them. In our validation sample, these are $1.2$--$2.7\times10^3$ gluon jets and $2.0$--$3.9\times10^3$ quark jets per $\ptjet$ bin, and the network correctly identifies $24\%$, $27\%$, $27\%$, $33\%$ and $34\%$ of the gluon jets and $20\%$, $22\%$, $28\%$, $31\%$ and $32\%$ of the quark jets, from the lowest to the highest $\ptjet$ bin. That these fractions increase with $\ptjet$ is consistent with the increasing gain of the network over the multiplicity in sec.~\ref{sec:auc}, and with the growing spread of $\Delta$ at fixed $N$. To illustrate that such fractions are expected even for an optimal classifier, app~\ref{app:bayes} constructs a toy scenario in which the exact likelihood ratio is known.

\subsection{Response to the radial spread of the jet}
\label{sec:radial}

Among the global observables of sec~\ref{sec:indep}, jet angularity $\lambda_1$ adds the most discriminating power to multiplicity, which makes the radial distribution of the jet momentum a natural first candidate for the content of $\Delta$. Instead of comparing $\Delta$ with a classifier built from $\lambda_1$, we deform the radial spread of each jet directly and study how the network output responds. To this end, we rescale the angular coordinates of all constituents relative to the jet axis as
\begin{equation}
\Delta\eta_i\to e^{\delta}\,\Delta\eta_i\,,\qquad\Delta\phi_i\to e^{\delta}\,\Delta\phi_i\,,
\label{eq:rescale}
\end{equation}
such that $r_i\to e^{\delta}r_i$, while keeping multiplicity, constituent momenta, charges and particle types fixed. 

Each angularity of eq.~\eqref{eq:obs} then changes as $\lambda_\beta\to e^{\beta\delta}\lambda_\beta$, i.e. $\delta>0$ makes the jet broader, as is typical of gluon jets, and $\delta<0$ narrower, as is typical of quark jets. Since $N$ and $\ptjet$ are unchanged, $\LN$ is unchanged as well, and any change of $M$ is a change of $\Delta$ alone. Moreover, a uniform rescaling preserves the ordering of the distances between constituents, so each constituent keeps the same nearest neighbors in the first EdgeConv block and only the angular separations themselves grow or shrink. We define the radial response of a jet to this deformation as
\begin{equation}
S_R=\frac{M(+\delta_0)-M(-\delta_0)}{2\delta_0}\simeq\frac{\partial\Delta}{\partial\delta}\,,\qquad\delta_0=0.05\,,
\label{eq:SR}
\end{equation}
which corresponds to a change of the width of the jet by about $5\%$. As a result, a value of $S_R>0$ would imply that the network considers a broader version of the same jet more gluon-like, as expected from the larger color charge of the gluon.

\begin{figure}[t]
    \centering
    \includegraphics[width=0.85\linewidth]{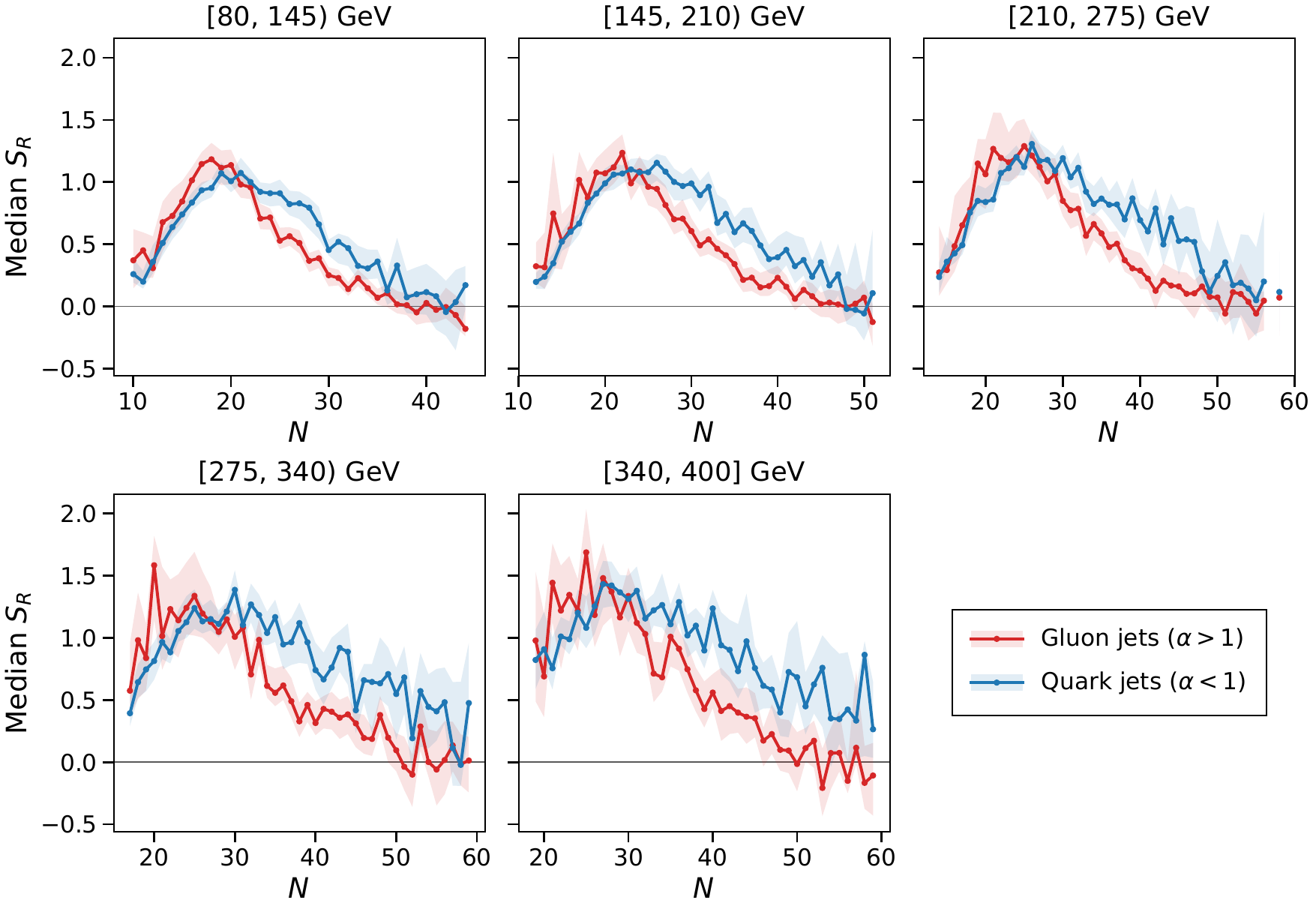}
    \caption{Median radial response $S_R$, eq.~\eqref{eq:SR} as a function of constituent multiplicity in five bins of $\ptjet$, for gluon jets (red) and quark jets (blue), after reweighting the jets at each $N$ to a common fine $(\ptjet,|y|)$ distribution. For quark jets, the response is evaluated under compression ($\alpha <1$), which gives the same derivative of $M$ with the opposite sign convention. Bands show 95\% bootstrap intervals .}
    \label{fig:SR1}
\end{figure}

\begin{figure}[t]
    \centering
    \includegraphics[width=0.85\linewidth]{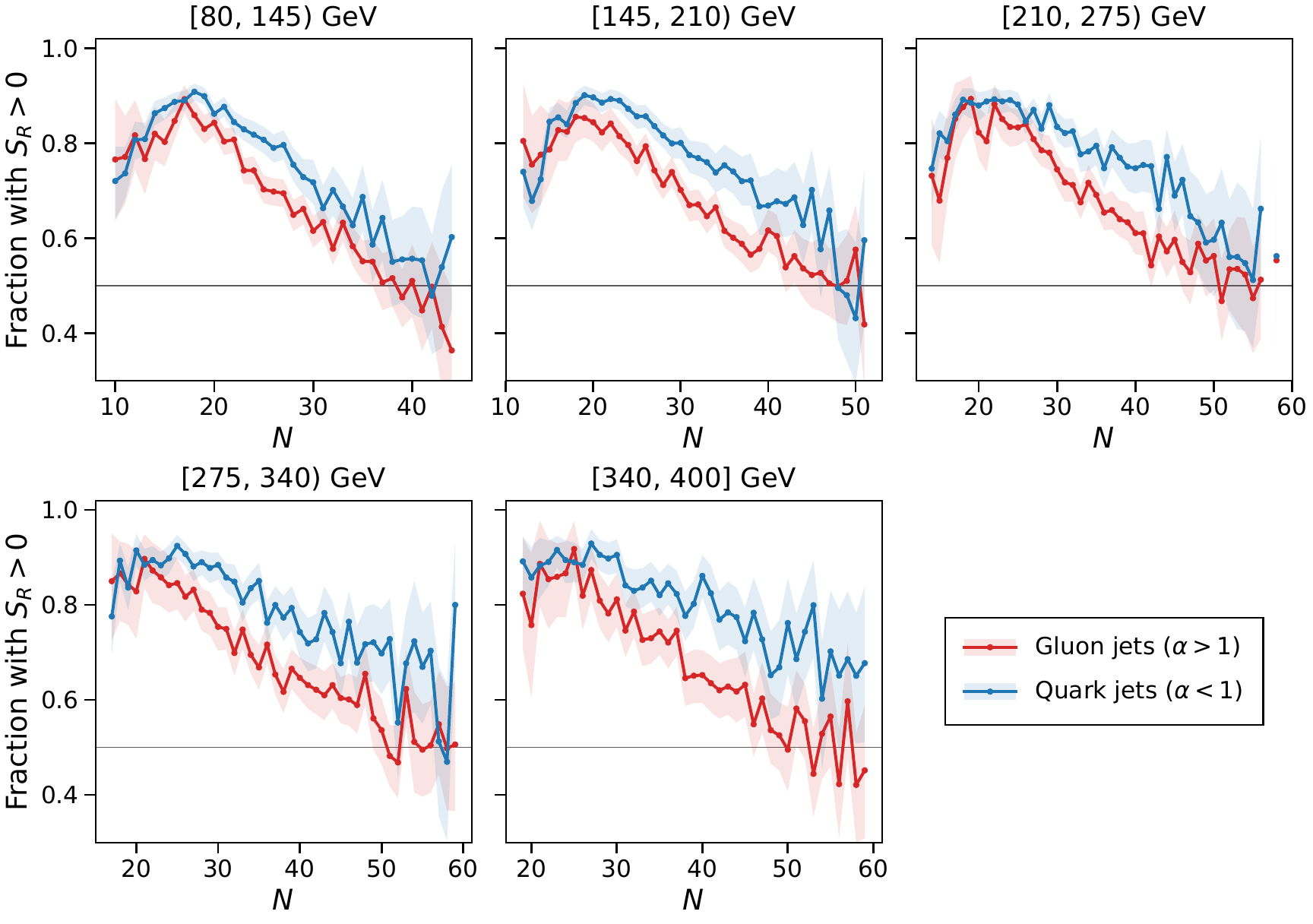}
    \caption{Same as fig.~\ref{fig:SR1} but now for fraction of jets with $S_R>0$.}
    \label{fig:SR2}
\end{figure}

\begin{figure}[t]
    \centering
    \includegraphics[width=0.85\linewidth]{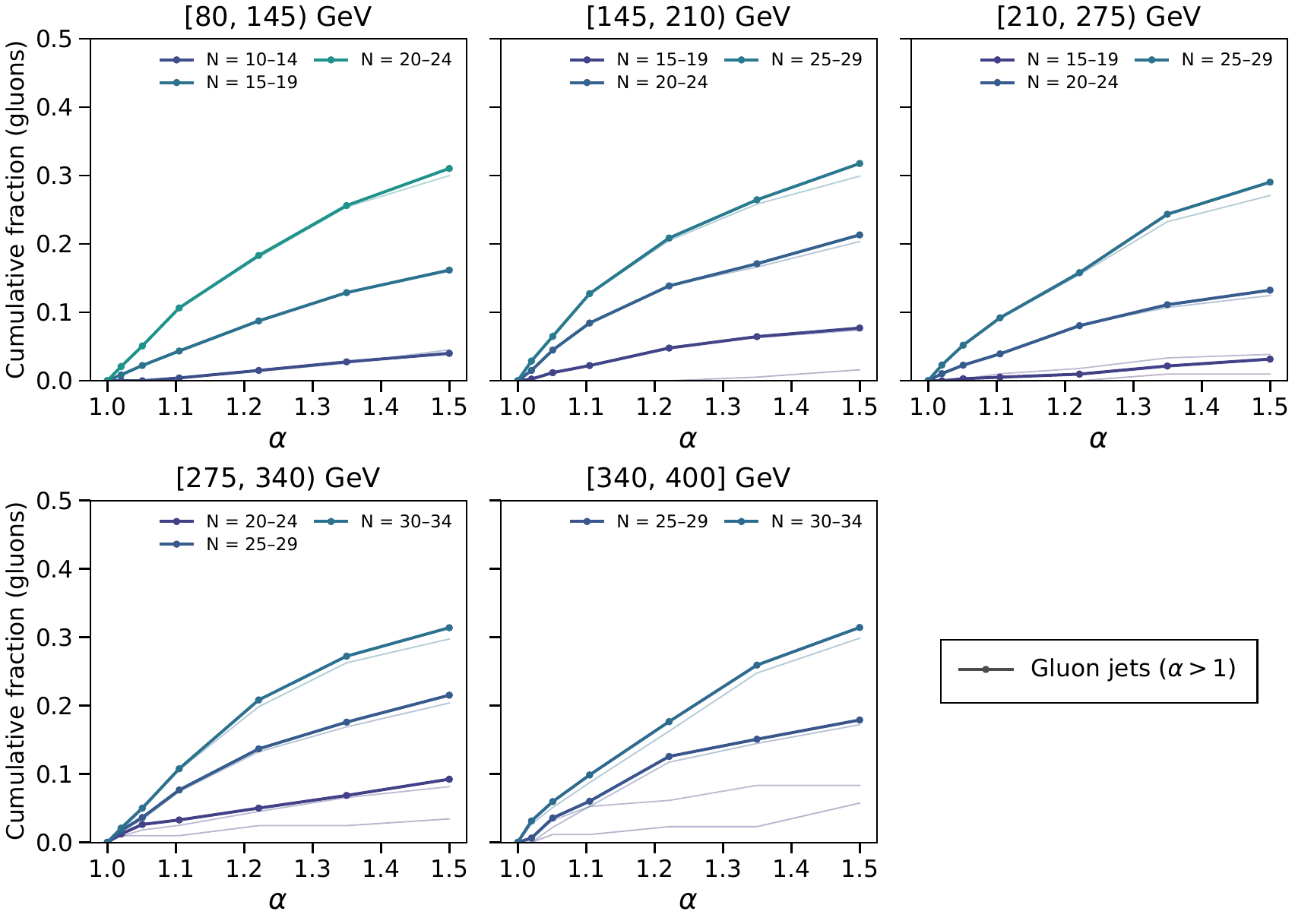}
    \caption{Cumulative fraction of gluon jets that change their predicted class as a function of the radial scale factor $\alpha=e^{\delta}$, ($\alpha>1$) in bins of low multiplicity, where misidentified gluon jets become correctly identified.}
    \label{fig:crossing1}
\end{figure}

\begin{figure}[t]
    \centering
    \includegraphics[width=0.85\linewidth]{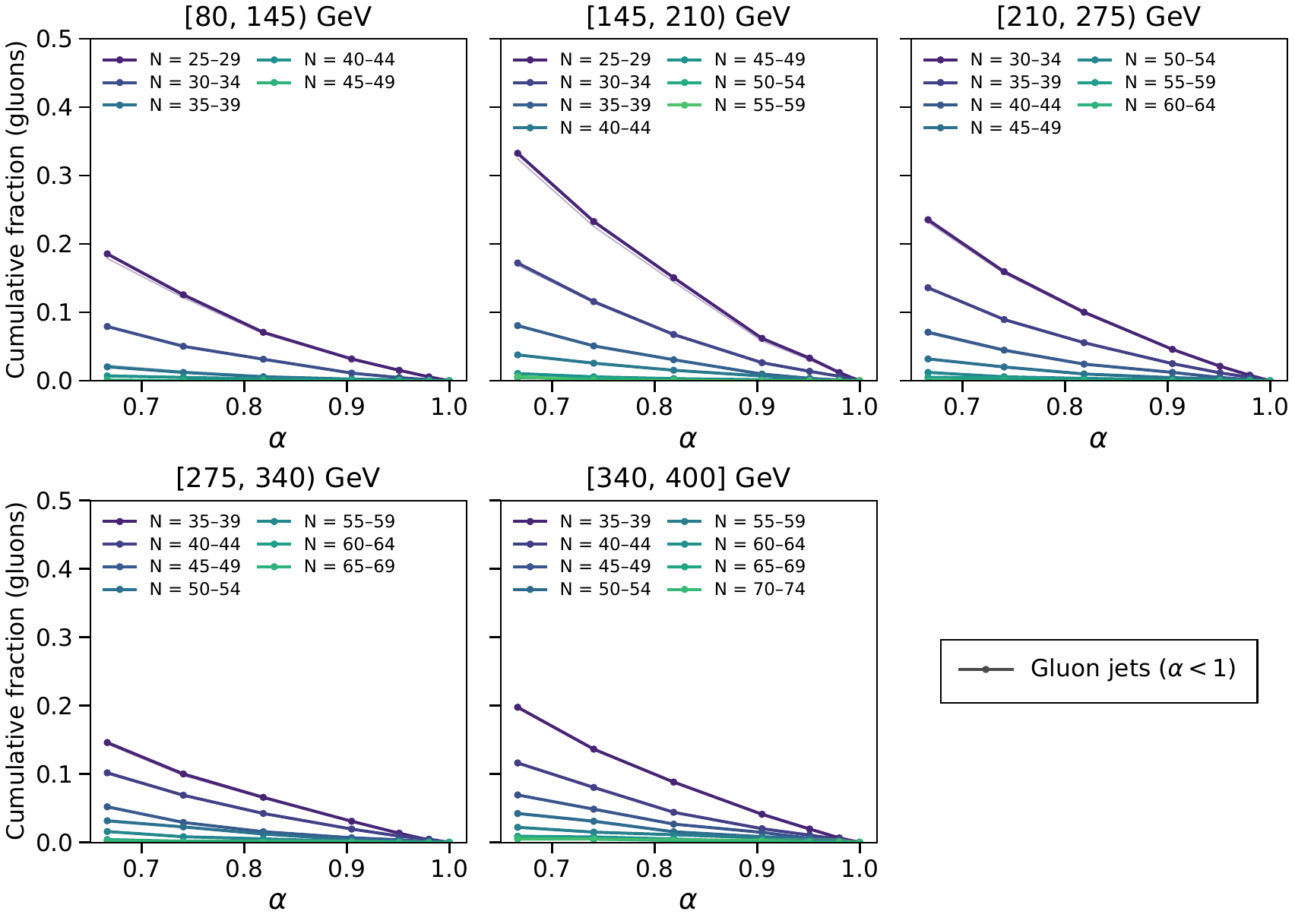}
    \caption{Same as fig.~\ref{fig:crossing1} but now $\alpha<1$ in bins of high multiplicity, where correctly identified gluon jets become misidentified.}
    \label{fig:crossing2}
\end{figure}

Figures~\ref{fig:SR1} and~\ref{fig:SR2} show the median radial response as a function of jet multiplicity in five $\ptjet$ bins. 
The response is positive for a large majority of jets. Moreover, the response of the network is largest at low and intermediate multiplicity, with a maximum at $N\simeq18$--$27$ depending on $\ptjet$, where $80$--$90\%$ of the jets become more gluon-like when broadened. 

Towards high multiplicity the response decreases, and for gluon jets it approaches zero in all $\ptjet$ bins, while for quark jets it remains positive, with a median of $0.3$--$0.5$ in the two highest bins. At low multiplicity, the quark and gluon curves are close. This is expected, since $S_R$ is the same derivative of $M$ for both classes, and at the same $N$ and $\ptjet$ the two can only differ through other properties of the jet substructure.

The radial spread thus matters most where the multiplicity alone is either quark-like or ambiguous. In other words, for a gluon jet with a low multiplicity it is mostly its broadness that can move it towards the gluon side, whereas a quark jet with a high multiplicity can still be identified as a quark jet if it is narrow. 
It is important to note, however, that gluon jets of high multiplicity already have large values of $M$, and the output of the network changes less for jets whose class is already well determined. Whether the decrease at high multiplicity reflects the physics of the jets or only their distance from the decision boundary would require a comparison at fixed $N$ and fixed $M$, which is beyond the scope of this work.

While $S_R$ describes the local response of each jet, figs.~\ref{fig:crossing1} and~\ref{fig:crossing2} show what happens for finite changes of the radial spread. Broadening gluon jets of low multiplicity by up to $50\%$ moves up to about $30\%$ of them to the gluon class when their multiplicity is such that both classes are equally likely, but less than $10\%$ at the lowest multiplicities, where $|\LN|$ is large. Correspondingly, narrowing gluon jets of high multiplicity by the same factor moves up to $15$--$33\%$ of them to the quark side close to that point, and almost none at the highest multiplicities. 
Thus, at fixed multiplicity, a realistic change of the radial spread can change the class of a jet only when its multiplicity does not strongly favor either class.

\begin{figure}[t]
    \centering
    \includegraphics[width=0.85\linewidth]{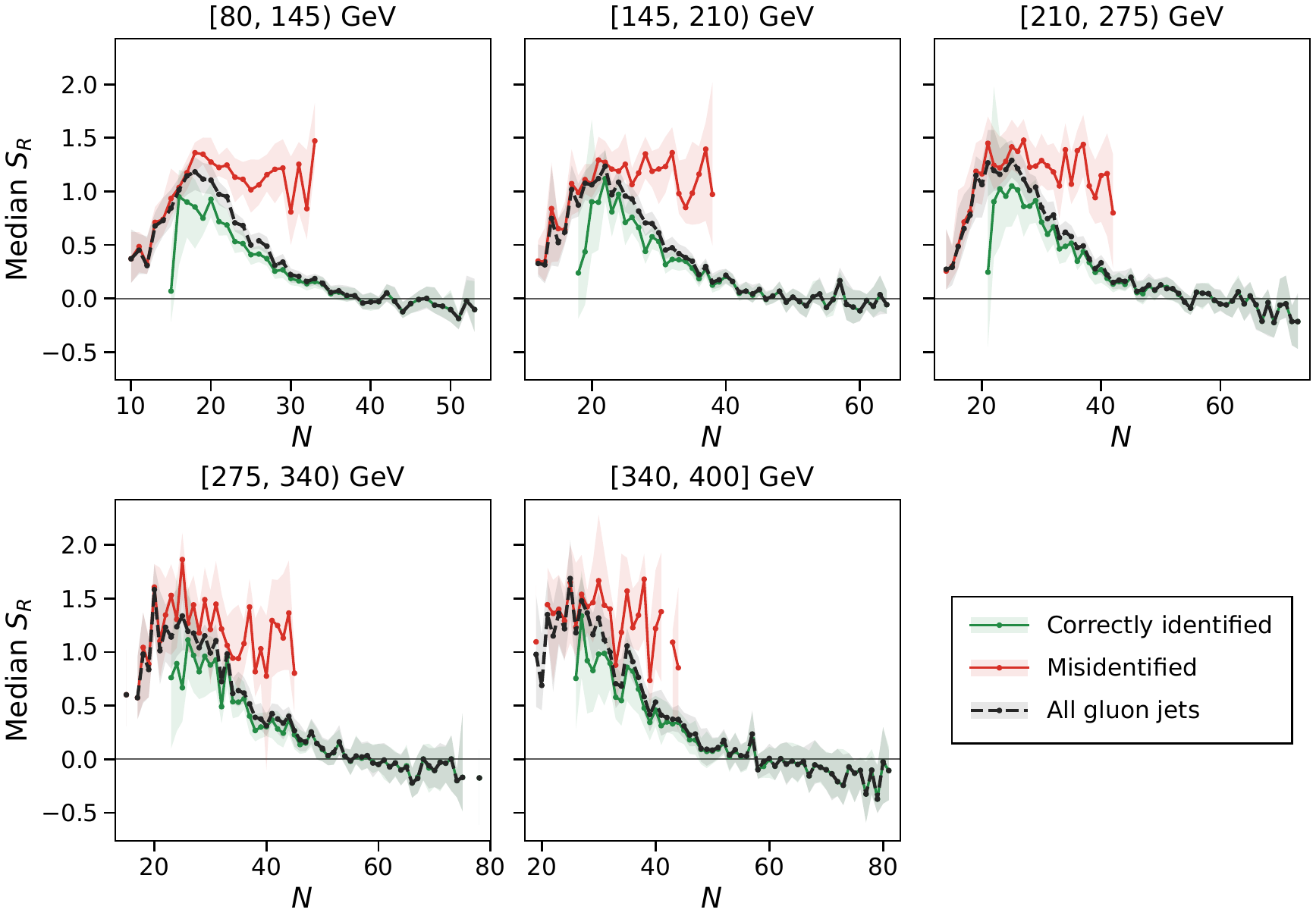}
    \caption{Median radial response $S_R$ as a function of constituent multiplicity for correctly identified (green), misidentified (red) and all (black) gluon jets in five bins of $\ptjet$. Bands show 95\% bootstrap intervals.}
    \label{fig:misID1}
\end{figure}

\begin{figure}[t]
    \centering
    \includegraphics[width=0.85\linewidth]{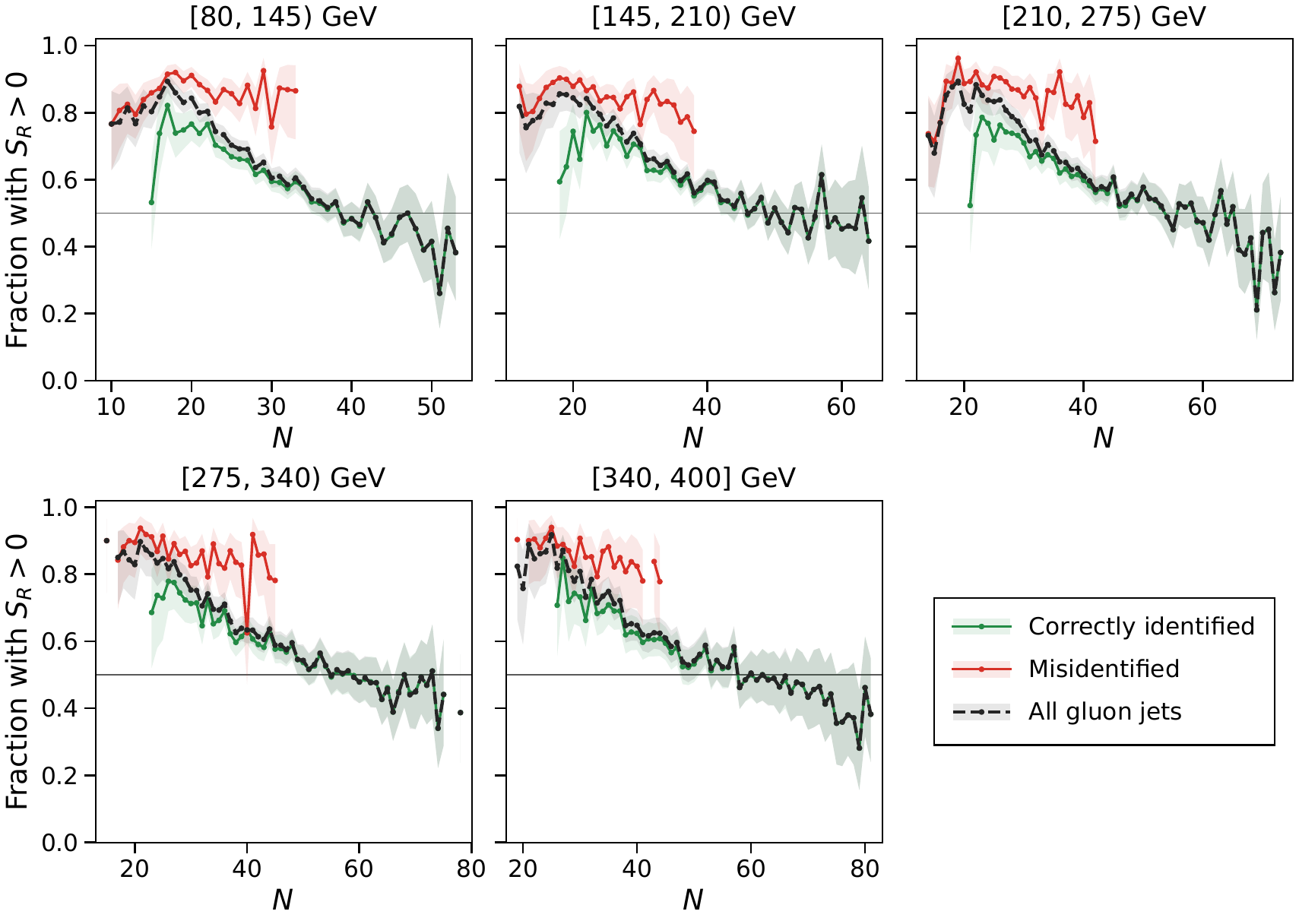}
    \caption{Same as fig.~\ref{fig:misID1} for fraction of jets with $S_R>0$.}
    \label{fig:misID2}
\end{figure}

Finally, figs.~\ref{fig:misID1} and~\ref{fig:misID2} compares the radial response of misidentified and correctly identified gluon jets of the same multiplicity. At low and intermediate multiplicity, misidentified gluon jets respond more strongly than correctly identified ones. This difference, however, must be interpreted with caution as the two populations are selected by the sign of $M$, i.e.\ by $\Delta$ itself, and therefore differ in any property that is correlated with $\Delta$, irrespective of the fact whether or not the network uses it.~\footnote{In the toy set-up of app~\ref{app:bayes}, where the classifier is exact and the same function for all jets, we show that misidentified gluon jets respond more or less strongly than correctly identified ones depending only on whether the gluon distribution of the second observable is narrower or wider than the quark one.} Figs.~\ref{fig:misID1} and~\ref{fig:misID2} therefore tell us where in the space of jets the misidentified gluon jets sit, i.e \ in the region where the network output is most sensitive to the radial spread consistent with narrow gluon jets close to the decision boundary.

\begin{figure}[t]
	\centering
	\includegraphics[width=0.95\linewidth]{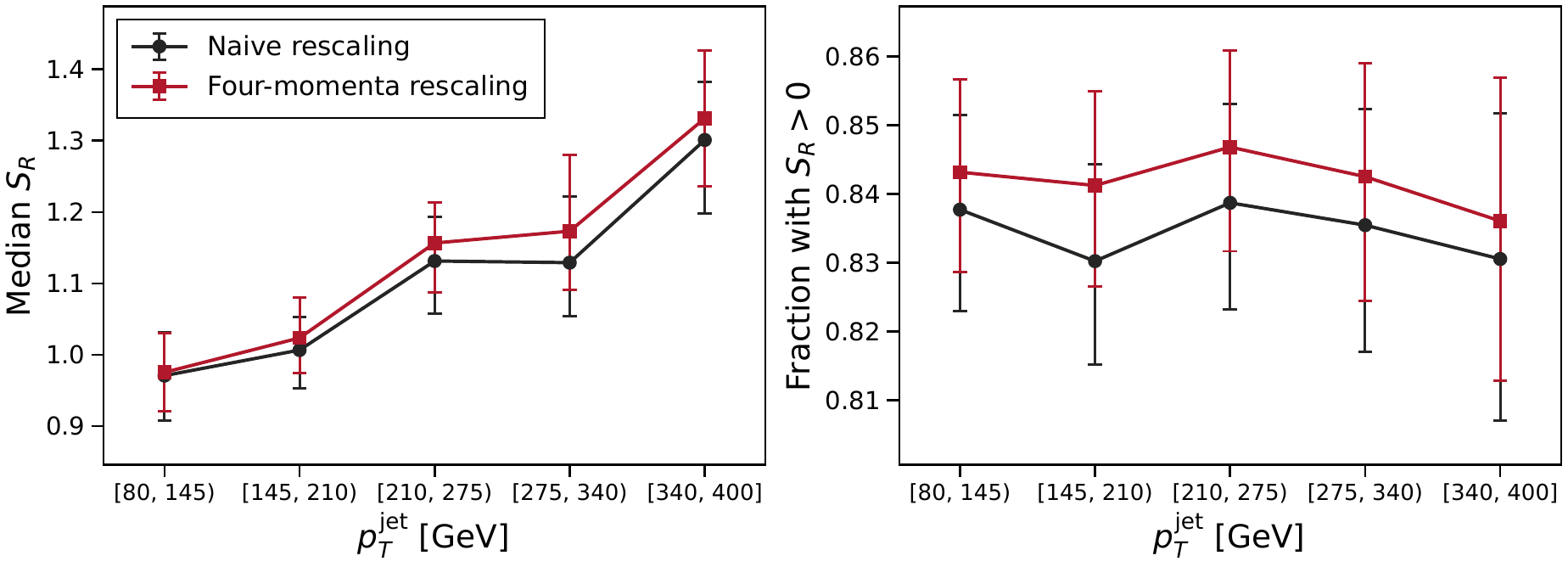}
	\caption{Comparison of median radial response and fraction of jets with $S_R > 0$ under a realistic modification of the particle four momenta with the naive expansion or compression of the radial components.}
	\label{fig:4-vec-reco}
\end{figure}

Next, to verify that the response is not an artifact of modifying the angular inputs without changing the energies of the constituents consistently, we repeat the rescaling of eq.~\eqref{eq:rescale} on the constituent four-vectors, combining the rescaled directions with the original transverse momenta and masses, recomputing the jet axis and all jet-level inputs, and recentring the constituents on the new axis. The result compared with our naive rescaling of the radial spread of the section is shown in fig.~\ref{fig:4-vec-reco}. From the figure, we see that the radial response of the network still survives under realistic modification of the jet radial distribution.

\section{Summary and outlook}
\label{sec:summary}

In this work, we studied how the discriminating power of a ParticleNet quark/gluon classifier is distributed between  constituent multiplicity and the remaining structure of the jet, differentially in jet transverse momentum and multiplicity. We trained the network on quark and gluon jets from $Z$+jet production in \textsc{Pythia}, matched in transverse momentum and rapidity over $80\le\ptjet\le400$~GeV, such that the network can only separate the two classes through the substructure of the jets. The network reaches an AUC that increases from $0.84$ to $0.90$ across this range.

To place this gain in context, we compared the network to classifiers built from standard jet observables. Using a boosted decision tree on six observables, we found that multiplicity, momentum dispersion and angularities carry largely shared information. Adding jet angularity and momentum dispersion to multiplicity provides a gain of about $2\%$ in  the AUC over multiplicity alone . However, at high $\ptjet$, multiplicity carries more discriminating information of its own. 

We then investigated whether the network output can be interpreted as a log-likelihood ratio. We found that, at fixed multiplicity, the mean gluon probability assigned by the network agrees well with the observed gluon fraction in all $\ptjet$ bins. However, jets of the same multiplicity nevertheless receive a broad range of gluon probabilities, i.e. the spread of network probability about the mean is large. This motivated us to decompose the network output as a sum of the multiplicity log-likelihood ratio and a residual log-likelihood ratio at fixed multiplicity. In this form, a jet whose multiplicity favors the opposite class is correctly identified only if the residual term overcomes the multiplicity bias. We find that this happens for $20$--$34\%$ of these jets, increasingly with $\ptjet$. The remaining jets reflect the overlap of quark and gluon jets, which even an ideal classifier would not be able to resolve. This decomposition of the network output as a multiplicity term and a residual term at fixed multiplicity could provide an interesting way to understand how quark/gluon taggers respond to quenched jets, which we will explore in a future study.

Finally, we studied to which property of the jet the residual term responds by rescaling the angular coordinates of the constituents at fixed multiplicity, momenta, charges and particle types. The network considers broader jets more gluon-like, and this response is largest at low and intermediate multiplicity, where the multiplicity alone does not decide the class of a jet. 

 \acknowledgments
AB and BS acknowledge the support by Funda\c{c}\~ao para a Ci\^encia e a Tecnologia (FCT) through the ERC-PT A-Projects `Unveiling', financed by PRR, NextGenerationEU. BS is also supported Fundação
para a Ciência e a Tecnologia (FCT) under contracts 2023.15319.PEX (https://doi.org/10.54499/2023.15319.PEX).

\appendix

\section{Construction of the multiplicity likelihood ratio}
\label{app:LN}

The multiplicity likelihood ratio of eq.~\eqref{eq:LN} requires the multiplicity distributions $p(N\,|\,c,\ptjet)$ of quark and gluon jets, $c=q,g$, as a function of the jet transverse momentum. We estimate them from the jets of the training sample only, such that neither the validation jets on which the network is evaluated nor the network itself enter this construction. The distributions are determined separately in each of the 64 fine $\ptjet$ bins $b$ of 5~GeV width used for the kinematic matching.

The histogram of $N$ by itself is not sufficient for this purpose as in the tails of the  distribution, only few jets populate and the histogram fluctuates strongly or is empty. As a result, the ratio in eq.~\eqref{eq:LN} becomes unstable or undefined. We therefore use a regularized histogram estimate. The histogram of each class is smoothed in $N$ with a Gaussian kernel of width $\sigma_N$, and a fixed number of pseudo-counts is added, which ensures that the estimated distributions are non-zero for all $N$ while having a negligible effect wherever a bin is well populated. The multiplicity likelihood ratio is then obtained as the logarithm of the ratio of the regularized gluon and quark distributions.

The smoothing width and the number of pseudo-counts are chosen from the training sample by five-fold cross validation: the training jets are divided into five parts, the distributions are estimated from four of them, and the log-likelihood of the multiplicities of the jets in the fifth part is evaluated. Maximizing this log-likelihood, averaged over the five choices of the held-out part, selects $\sigma_N=1.5$ and 30 pseudo-counts. The resulting $\LN$ depends on $\ptjet$ only, and the same values are used for all rapidities.

Finally, the network receives at most the 100 highest-$\pT$ constituents of each jet, such that a stored multiplicity of $N=100$ only means that the jet has at least 100 constituents. Since $\LN$ is not defined for such jets, the 19 jets of the validation sample with $N=100$ are not used in the analyses of sections~\ref{sec:prob} and~\ref{sec:beyond}.

\section{Bayes benchmark}
\label{app:bayes}

\begin{figure}[!htb]
	\centering
	\includegraphics[width=0.48\textwidth]{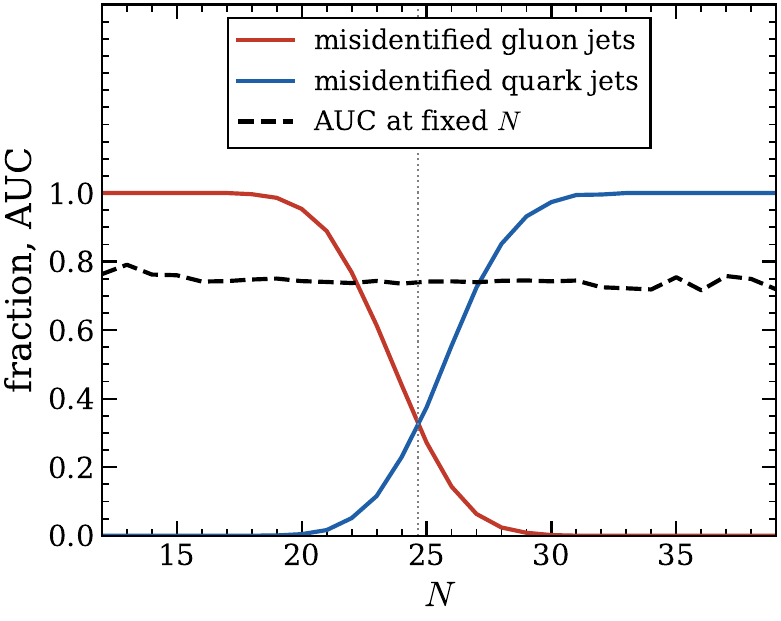}\\[2mm]
	\includegraphics[width=0.96\textwidth]{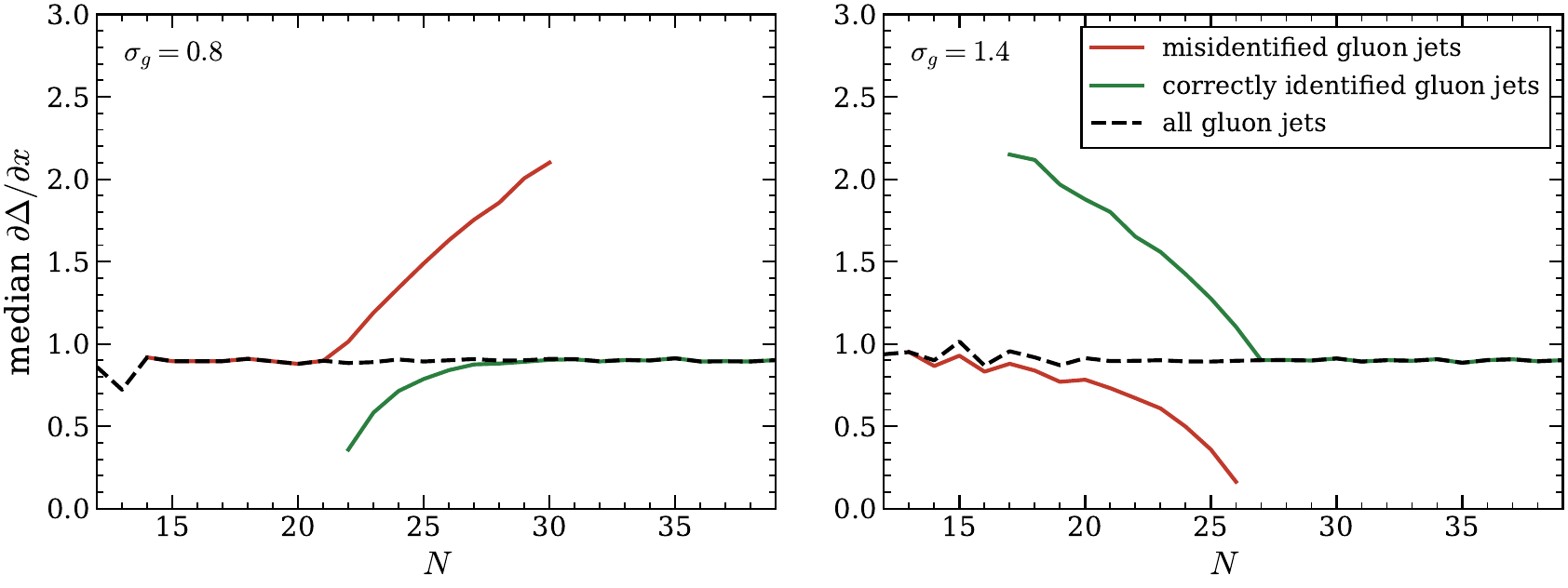}
	\caption{Fractions of misidentified gluon and quark jets for an exact classifier as a function of $N$, and its AUC at fixed $N$, for $\sigma_g=1$ (top). The dotted line marks the multiplicity at which both classes are equally likely. Median response $\partial\Delta/\partial x$ of misidentified, correctly identified and all gluon jets as a function of $N$, for a gluon distribution of $x$ narrower ($\sigma_g=0.8$, left) and wider ($\sigma_g=1.4$, right) than the quark one (bottom).}
	\label{fig:bayes}
\end{figure}

In this appendix, we employ a toy set-up elaborating why a comparison between misidentified and correctly identified jet classes should be interpreted with care.  For this purpose, we  construct a classifier that is exact and hence optimal by construction. In this set-up, each jet is described by two quantities. First is its multiplicity, which we assume to follow a Poisson distribution with mean $\mu_q=20$ for quark and $\mu_g=30$ for gluon jets, mimicking the fact that gluon jets have larger multiplicities. The second parameter our toy model is sensitive to represents a residual property of the jet at fixed multiplicity, for instance its radial spread. We again take a simple ansatz for this, for instance, a Gaussian with mean $0$ and unit width for quark jets and mean $0.9$ and width $\sigma_g$ for gluon jets, such that this property is independent of $N$ for a given class. The exact log-likelihood ratio then splits as in eq.~\eqref{eq:decomp}, into a multiplicity term $\LN$ and a residual term $\Delta(x)$, which is quadratic in $x$, and linear if $\sigma_g=1$.

We first consider $\sigma_g=1$. Figure~\ref{fig:bayes} (top) shows that the fraction of misidentified gluon jets falls and that of misidentified quark jets rises with $N$, crossing at multiplicity values where both classes are equally likely, while the AUC at fixed $N$ stays at $0.72$--$0.79$. Of the jets whose multiplicity favors the opposite class by odds of at least $2{:}1$, the exact classifier correctly identifies only $11\%$ of the gluon jets and $14\%$ of the quark jets. Note that these numbers do not carry any physical significance and depend on the parameters of the toy set-up but are meant to show that even for an exact classifier, the recovery of the misidentified jets is mostly limited by the genuine overlap between the underlying distributions.

We then compute the response $\partial\Delta/\partial x$ of our exact classifier and compare misidentified and correctly identified gluon jets of the same multiplicity, as in figs.~\ref{fig:misID1} and~\ref{fig:misID2}. We do this for a gluon distribution of $x$ with $\sigma_g=0.8$ and  $\sigma_g=1.4$. Figure~\ref{fig:bayes} (bottom) shows that the two groups differ in both cases, but in opposite directions: misidentified gluon jets have a median response of $1.01$ compared to $0.89$ for correctly identified ones in the first case, and $0.78$ compared to $0.91$ in the second. This difference arises because the two groups are selected by the value of $\Delta$ and therefore sit in different regions of $x$, where the slope of $\Delta$ differs. A difference between misidentified and correctly identified jets at fixed multiplicity therefore tells us where the two groups sit in the classifier's decision boundary rather than on how the classifier uses the feature $x$.

\bibliographystyle{utphys.bst}
\bibliography{refs.bib}

\end{document}